\documentclass[superscriptaddress,prd,showpacs,twocolumn,showpacs,nofootinbib]{revtex4-2}
\usepackage{eurosym}
\usepackage{graphicx}
\usepackage{color}
\usepackage{epsf}
\usepackage{bm}
\usepackage{amsmath}
\usepackage{amsfonts}
\usepackage{amssymb}
\usepackage{epstopdf}
\usepackage{multirow}
\usepackage{hhline}
\usepackage[dvipsnames]{xcolor}

\newcommand{\lsim}   {\mathrel{\mathop{\kern 0pt \rlap
  {\raise.2ex\hbox{$<$}}}
  \lower.9ex\hbox{\kern-.190em $\sim$}}}
\newcommand{\gsim}   {\mathrel{\mathop{\kern 0pt \rlap
  {\raise.2ex\hbox{$>$}}}
  \lower.9ex\hbox{\kern-.190em $\sim$}}}
\newcommand{\bw}{\begin{widetext}\begin{equation}}
\newcommand{\ew}{\end{equation}\end{widetext}}
\newcommand{\be}{\begin{equation}}
\newcommand{\ee}{\end{equation}}
\newcommand{\bea}{\begin{eqnarray}}
\newcommand{\eea}{\end{eqnarray}}

\begin{document}

\title{Static spherically symmetric three-form stars}

\author{Bruno J. Barros}
\email{bjbarros@fc.ul.pt}
 \affiliation{Instituto de Astrofísica e Ci\^encias do Espa\c{c}o, Faculdade de Ci\^encias da Universidade de Lisboa, Edif\'icio C8, Campo Grande, P-1749-016, Lisbon, Portugal}

\author{Zahra Haghani}
\email{z.haghani@du.ac.ir}
  \affiliation{School of Physics, Damghan University, Damghan, 41167-36716, Iran,}

 \author{Tiberiu Harko}
 \email{tiberiu.harko@aira.astro.ro}
 \affiliation{Astronomical Observatory, 19 Ciresilor Street, 400487 Cluj-Napoca, Romania}
\affiliation{Department of Physics, Babes-Bolyai University, Kogalniceanu Street,
Cluj-Napoca 400084, Romania,}
\affiliation{School of Physics, Sun Yat-Sen University, Guangzhou 510275, People's Republic of China}
\author{Francisco S. N. Lobo}
 \email{fslobo@fc.ul.pt}
 \affiliation{Instituto de Astrofísica e Ci\^encias do Espa\c{c}o, Faculdade de Ci\^encias da Universidade de Lisboa, Edif\'icio C8, Campo Grande, P-1749-016, Lisbon, Portugal}

\date{\today}

\begin{abstract}
We consider interior static and spherically symmetric solutions in a gravity
theory that extends the standard Hilbert-Einstein action with a Lagrangian
constructed from a three-form field $A_{\alpha \beta \gamma}$, which
generates, via the field strength and a potential term, a new component in
the total energy-momentum tensor of the gravitational system. We formulate
the field equations in Schwarzschild coordinates and investigate their
solutions numerically for different equations of state of neutron and quark
matter, by assuming that the three field potential is either a constant or
possesses a Higgs-like form. Moreover, stellar models, described by the stiff
fluid, radiation-like, bag model and the Bose-Einstein condensate equations
of state are explicitly obtained in both general relativity and three-form
gravity, thus allowing an in-depth comparison between the astrophysical
predictions of these two gravitational theories. As a general result we find
that for all the considered equations of state, three-form field stars are
more massive than their general relativistic counterparts. As a possible
astrophysical application, we suggest that the 2.5$%
M_{\odot}$ mass compact object, associated with the GW190814 gravitational
wave event, could be in fact a neutron or a quark star described by the
three-form gravity theory.
\end{abstract}


\maketitle
\tableofcontents





\section{Introduction}\label{introduction}

The experimental detection of gravitational waves by the LIGO and VIRGO
scientific collaborations \cite{Abbott:2016blz,TheLIGOScientific:2016wfe} has opened a new window on the
intricate physical processes that control gravitational phenomena,
leading to a better understanding of the properties of compact objects. The
important GW170817 event \cite{TheLIGOScientific:2017qsa} initiated the Multimessenger Era,
with the signal, originating from the shell elliptical galaxy NGC 4993, and
produced by the merging of two neutron stars, detected by
more than 60 instruments worldwide.  GW170817  suggests a constraint on the mass of the nonrotating neutrons star given by $M\leq 2.3M_{\odot}$. For a review of the merger of binary
neutron star systems we refer the reader to \cite{Baiotti:2016qnr}. The merger of neutron stars that takes
place in the conditions of extreme gravity leads to the emission of intense
fluxes of gravitational waves, associated with complicated microphysical and
electromagnetic processes. The resulting astrophysical signatures can be
observable even at the highest redshifts.

The detection of gravitational
waves has also lead to some observations that are likely to change some
basic paradigms in astrophysics. One such observation is related to the
GW190814 event \cite{Abbott:2020khf}, which has shown a very intriguing structure of the
mass components, with one of the individual masses in the range $2.5 - 2.67M_{\odot}$
(90\% confidence). No optical counterpart to the gravitational wave was
observed. If one assumes that this object is a neutron star, its high mass
would strongly contradict the paradigm of the existence of a standard $%
1.4M_{\odot}$ mass scale for compact stars. In fact, a Bayesian statistical
inference, performed in \cite{Valentim:2011vs}, evaluating the likelihood of the proposed
Gaussian peaks by using 54 measured points obtained in a variety of systems,
has already concluded on the existence of a bimodal distribution of the
masses, with the first peak around 1.37 $M_{\odot}$, and a much wider second
peak at 1.73 $M_{\odot}$.

However, the mass observed in the GW190814 event
is much higher than these peaks. On the other hand, another accurate
measurement of the mass of a compact object, using Shapiro delay, yielded
the value $2.14 ^{+ 0.10}_{ -0.09}M_{\odot}$ for the mass of the millisecond
pulsar MSP J0740+6620 \cite{Cromartie:2019kug}. A few higher measured mass values also
exist. These observations raise the question of the maximum mass $%
M_{\rm max}$ of stable compact general relativistic objects, since in order to
accommodate the observed values one must significantly enlarge the allowed
range of $M_{\rm max}$. In turn, this would require important modifications for
the physical properties of the dense matter at densities higher than the
saturation density, and a corresponding modification of the equation of
state. Even more interesting, and with important theoretical consequences,  is the recent discovery of a companion, having a mass of around $3M_{\odot}$, of V723 Mon, a nearby evolved red giant in a high mass function, $f(M)=1.72\pm 0.01M_{\odot}$, nearly circular binary \cite{BHm}. If confirmed, this discovery will raise new questions about the present day knowledge of the mass distribution of massive compact objects, and on the transition of neutron stars to black holes.

There are a number of effects that could lead to the increase of the masses
of compact stars. One such effect is related to the assumption of a phase
transition in the dense matter, leading to the formation of a quark star. In
\cite{Kovacs:2009kv} the possibility that stellar mass black holes, with masses in the
range of 3.8$M_{\odot}$ and 6$M_{\odot}$, could be in fact quark stars in
the Color-Flavor-Locked (CFL) phase was investigated in detail. It was shown
that, depending on the value of the gap parameter, rapidly rotating CFL
quark stars may have much higher masses than ordinary neutron stars. On the
other hand quark stars have a very low luminosity, and an almost completely
absorbing surface, due to the fact that infalling matter on the surface of
the quark star is fully transformed into quark matter.

A possibility of
distinguishing quark stars in standard or CFL phase from low mass black
holes or neutron stars could be through the study of thin accretion disks
around rapidly rotating stars (neutron or quark), and Kerr type black holes,
respectively. It was already suggested that the GW190814 event resulted from
the merging of a black hole -- strange quark star system \cite{Bombaci:2020vgw,Wu:2020zhr,Horvath:2020lwj}.
Some compact astrophysical objects may contain a significant part of their
matter in the form of a Bose-Einstein condensate. The basic astrophysical
parameters (mass and radius) of the neutron stars sensitively depend on the
mass of the condensed particle, and on the scattering length. Hence the
recently observed neutron stars with masses in the range of 2-2.5 $M_{\odot}$
could be Bose-Einstein Condensate stars, containing a large amount of
superfluid matter \cite{Chavanis:2011cz}.

An alternative explanation of the higher masses of some classes of neutron
stars is related to the possible modification of the very nature of the
gravitational force at very high densities, implying that the structure of
neutron stars is described by some modified theories of gravity. In the
presence of a cosmological constant $\Lambda$ the mass-radius $M/R$ ratio of
compact objects satisfy the constraint $2M/R\leq \left(1-8\pi \Lambda R^2/3
\right)\left[1-\left(1-2\Lambda/\bar{\rho}\right)^2/9\left(1-8\pi \Lambda
R^2/3\right)\right]$ \cite{Mak:2001gg}. Upper and lower bounds on the mass-radius
ratio of stable compact objects in extended gravity theories, in which
modifications of the gravitational dynamics with respect to standard general
relativity are described by an effective contribution to the matter
energy-momentum tensor $T_{\mu}^{\nu}$, given by a tensor $\theta
_{\mu}^{\nu}$, were derived in \cite{Burikham:2016cwz}. By introducing the effective
density inside the star defined as $\rho _{\mathrm{eff}}c^{2} = \rho c^{2}/G
+ \theta _{0}^{0}$, and the effective mass $m_{\mathrm{eff}}=4\pi \int _0^r{%
r^2\rho _{\mathrm{eff}}dr}$, one obtains for the generalized Buchdahl limit
for extended gravitational theories the expression
\begin{equation}  \label{58}
\frac{2m_{\mathrm{eff}}(r)}{r}\leq 1-\left[1+\frac{2\left(1+f(r)\right)}{%
1+4\pi w_{\mathrm{eff}}(r)}\right]^{-2},
\end{equation}
where
\begin{equation}  \label{f(r)*}
f(r) = 4\pi \frac{\Delta (r)}{\langle \rho_{\mathrm{eff}}\rangle (r)}\left\{%
\frac{\arcsin\left [\sqrt{2m_{\mathrm{eff}}(r)/r}\right ]}{\sqrt{%
2m_{\mathrm{eff}}(r)/r}} -1\right \},
\end{equation}
$w_{\mathrm{eff}}(r)=p_{\mathrm{eff}}/\langle \rho_{\mathrm{eff}}\rangle (r)$%
, and $\Delta =\left(G/c^4\right)\left(\theta _1^1-\theta _2^2\right)$,
respectively. Hence, the extra contributions to the matter energy-momentum
tensor due to the modifications of the gravitational force could lead to a
significant increase in the mass of the neutron star.

 The structure and physical properties of specific classes of neutron, quark and ``exotic'' stars in Eddington-inspired Born-Infeld gravity were considered in \cite{Harko:2013wka}. The latter reduces to standard general relativity in vacuum, but presents a different behavior of the gravitational field in the presence of matter. The equilibrium equations for a spherically symmetric configuration (mass continuity and Tolman-Oppenheimer-Volkoff) were derived, and their solutions were obtained numerically for different equations of state of neutron and quark matter.
The internal structure and the physical properties of specific classes of neutron, quark and Bose-Einstein Condensate stars in the hybrid metric-Palatini gravity theory \cite{Capozziello:2013uya,Harko:2011nh,Capozziello:2012ny,Capozziello:2015lza,Harko:2020ibn}, which is a combination of the metric and Palatini $f(R)$ formalisms, was considered in \cite{Danila:2016lqx}. As a general result it was found that for all the considered equations of state, hybrid metric-Palatini gravity stars are more massive than their general relativistic counterparts. The properties of neutron stars in $f(R,T)$ gravity \cite{Harko:2011kv} for the case  $R+2\lambda T$,  where $R$ is the Ricci scalar and  $T$  is the trace of the energy-momentum tensor were investigated in \cite{Lobato:2020fxt}.  The hydrostatic equilibrium equations have been solved by considering realistic equations of state. It was also found that using several relativistic and non-relativistic models the variation on the mass and radius of the neutron star is almost the same for all considered equations of state, indicating that the results are independent of the high density part of the equation of state. Hence the stellar masses and radii depend only on the crust, where the equation of state is essentially the same for all the models.

The above-mentioned modified theories of gravity can be represented in a respective scalar-tensor theory. In fact, scalar fields in cosmology and gravitation have unquestionably played a crucial role over the last decades \cite{Clifton:2011jh,Capozziello:2011et}. Nonetheless, so far we have only detected one scalar particle in nature, namely, the Higgs boson \cite{Aad:2012tfa}. On a smooth manifold $\mathcal{M}$, scalar fields are part of a more general class of fields, denoted as $n$-forms, with scalars being the $n=0$ case. These differential forms inhabit smooth sections of the $n$-th exterior power of the cotangent bundle ${T^* \mathcal{M}\overset{\pi}{\longrightarrow} \mathcal{M}}$, and thus naturally exist in most geometrical settings. When performing calculus on manifolds, these fields have an enormous importance, since their algebraic structure allows us to construct diffeomorphism invariant objects and carry out integration in a coordinate-free fashion. Pioneered by \'{E}lie Cartan in the beginning of the 20th century, they are also the central objects of De Rham cohomology, linking the topological and differentiable structures of manifolds.
Since the geometrical anatomy of a manifold is intimately connected with the dynamical behaviour of theories defined within it, it is reasonable to assume that differential forms play a fundamental role in physics, particularly while formulating (gauge) field theories in a form-representative framework. Therefore it becomes relevant to explore the effects of considering higher spin fields in gravitation and cosmology.

Here, we will focus on three-form fields \cite{Koivisto:2009ew}. Three-forms naturally exist in fundamental theories, such as string theory and supergravity \cite{Connes:1997cr,Kachru:2003aw,Hoare:2013pma,Farakos:2017ocw,Ovrut:1997ur}, so it is not unreasonable to expect their emergence in low energy effective actions. Their first link to cosmology may be traced back to Hawking and Turok \cite{Turok:1998he}, when trying to explain the tiny value of the cosmological constant and thus realizing that an action encompassing a four-form, constructed from a three-form gauge field, behaves exactly as a cosmological constant. Ten years later, models of three-forms with the addition of self-interacting potentials were studied, where it was shown that these may give rise to self-accelerating attractor solutions \cite{Koivisto:2009fb}, useful to explain primordial inflation \cite{Kumar:2014oka,DeFelice:2012jt,Koivisto:2009sd,Barros:2015evi} and dark energy \cite{Koivisto:2012xm,Morais:2016bev,Wongjun:2016tva}. These models also present distinct observational signatures, so in principle, it would be possible to distinguish between three-form and standard scalar driven models \cite{Mulryne:2012ax}.

Three-forms were further applied to other scenarios, such as a mechanism for magnetogenesis \cite{Koivisto:2011rm} considering $U(1)$ couplings to the electromagnetic field, and three-form screening mechanisms around dense objects \cite{Barreiro:2016aln}. Three-forms cosmological models were analyzed in detail in \cite{new1}, where it was shown that in this kind of models there are fixed points at infinity. This result was  obtained by introducing appropriate compactifications, as well as by defining a new time variable that eliminates any potential divergence of the system. Normally hyperbolic non-isolated fixed points were also identified.

The employment of three-form fields in gravitation was analyzed in \cite{Barros:2018lca}, where wormhole solutions in a static and spherically symmetric spacetime continuum were explored. It was found that it is possible for the ordinary matter fields threading the wormhole to obey the classical energy conditions throughout the spacetime, while the three-form field holds the wormhole open, violating the null and weak energy conditions. It is known \cite{Koivisto:2009sd} that a three-form admits a dual scalar field representation. However, it is important to notice that this mapping easily breaks down when considering even fairly simple self interactions, nonminimal couplings, or noncanonical kinetic terms for the three-form \cite{Koivisto:2009fb}. When this dual representation is well defined, the scalar representation is often complicated and untractable. Hence the three-form formalism may provide for new and rich frameworks to explore physical phenomena. For a model of three-form fields in a homogeneous and isotropic universe a canonical quantization procedure of the Wheeler-DeWitt type was discussed in \cite{new2}, by obtaining first the Hamiltonian description of the model. This formalism was applied to a Little Sibling of the Big Rip (LSBR), an event that is known to appear in several minimally coupled three-form fields for a variety of potentials. A set of analytical solutions of the Wheeler-DeWitt equation was obtained, their physical implications were discussed. It was also shown that there are quantum states where the wave function of the universe vanishes, i.e. the DeWitt condition is fulfilled for them.

The static and spherically symmetric vacuum solutions in the three-form field gravity theory were also investigated in \cite{Barros:2020ghz}.  For the case of the vanishing three-form field potential the gravitational field equations can be solved exactly. However, for arbitrary potentials, due to their mathematical complexity, numerical approaches were adopted for studying the behavior of the metric functions and the three-form field. The formation of a black hole was detected from the presence of a Killing horizon for the time-like Killing vector in the metric tensor components. Several models, corresponding to different functional forms of the three-field potential, namely, the Higgs and exponential type, were considered. In particular, naked singularity solutions were also obtained for the exponential potential case. The thermodynamic properties of the black hole solutions, such as the horizon temperature, specific heat, entropy and evaporation time due to the Hawking luminosity, were investigated in detail. In \cite{new3} it was shown that a minimally coupled 3-form endowed with a proper potential can support a regular black hole interior,  and that by choosing an appropriate form for the potential of the 3-form field, one can construct an interior black hole spacetime that is everywhere regular. Moreover, the singularity is replaced with a Nariai-type spacetime, whose topology is $dS_2\times S_2$. Despite the negative potential of the 3-form field, the 2-dimensional de Sitter geometry always appears. Due to the violation of the null energy condition this particular geometry is singularity-free.

As an extension of the above-mentioned work, it is the goal of the present paper to consider the properties of compact high density stars in the three-form field gravity theory. To simplify the mathematical formalism we adopt a spherically symmetric static geometry, and a matter source. After writing down the gravitational field equations in their full generality, as a first step in our study we obtain the generalized mass continuity equation, the Tolman-Oppenheimer-Volkoff equation, describing the hydrostatic equilibrium of the star, as well as the field equation describing the variation inside the compact object of the non-zero component of the three-form field. These three equations fully describe the macroscopic
properties of the three-form field stars. The system of the structure equations of the three-form field gravity theory is then solved numerically for
several equations of state of the high density matter.

In our study we
consider three-form field stars whose ordinary baryonic matter content is described by the stiff fluid (Zeldovich) equation of state, satisfying the causality condition requiring that the speed of
sound in the dense matter does not exceed the speed of light, by the photon gas equation of state, describing a radiation fluid with the property,
that the trace of the energy-momentum tensor identically vanishes; the strange quark matter equation of state, derived from the MIT bag model, and, the equation of state of the Bose-Einstein Condensates, given by a polytropic equation of state with polytropic index
$n = 1$. For all these equations of state of the dense matter the astrophysical parameters of the stars (density, pressure and mass distribution, radius and total mass), as well as the behavior of the three-form field and of its potential are obtained numerically, and compared with the results of the similar standard general relativistic models. Hence this strategy allows us to perform an in-depth comparison of the two gravitational theories, and of the impact of modified gravity on the description of the properties of the stellar structures. We can formulate a general conclusion
of our study by pointing out that three-form field gravity theory predicts the existence of more massive high density stars, as compared to standard general relativity.

The present paper is organized as follows. The action and the field equations of the three-form field gravity theory is briefly introduced in Sec.~\ref{AaF}. The system of gravitational field equations, describing the interior of static spherically symmetric stars, are written down in Sec.~\ref{back}, where the structure equations of compact objects (mass continuity, Tolman-Oppenheimer-Volkoff, and three-form field evolution equation) are also derived, and rewritten in a dimensionless form. The global astrophysical properties of the Zeldovich (stiff fluid), photon, strange matter, and Bose-Einstein Condensate stars are obtained, by numerically integrating the structure
equations of the static spherically symmetric three-form field gravity theory, in Sec.~\ref{struct}.  We discuss and conclude our results in Sec.~\ref{conclusions}.

\section{Action and field equations}\label{AaF}

Let us start by considering a single canonical three-form field ${\bf A}$, with action given by \cite{Koivisto:2009fb,Koivisto:2009ew}
\begin{equation}
\label{action3F}
\mathcal{S}_A = - \int \omega_g\left[\frac{1}{48}F^{\alpha\beta\gamma\delta}F_{\alpha\beta\gamma\delta} + V(A^{\alpha\beta\gamma} A_{\alpha\beta\gamma}) \right],
\end{equation}
where $\omega_g = \sqrt{-g}\,{\rm d}^4x$ is the metric volume form, with $g$ being the determinant of the metric, and we have assumed a self-interacting potential $V$, which is a function of the invariant $A^{\alpha\beta\gamma} A_{\alpha\beta\gamma}=:A^2\,$. The first kinetic term in Eq.~\eqref{action3F},
\begin{equation}
-\frac{1}{48} \int \omega_g F^{\alpha\beta\gamma\delta}F_{\alpha\beta\gamma\delta} = -\frac{1}{2} \int \bf{F}\wedge \star\bf{F},
\end{equation}
comprises the 4-form strength tensor ${\bf{F}=d\bf{A}\,}$ \cite{Morais:2016bev,Mulryne:2012ax}, which generalizes Maxwell's 2-form from classical electromagnetism, with components ${F_{\alpha\beta\gamma\delta} = 4\nabla_{[\alpha}A_{\beta\gamma\delta ]}\,}$, and it is a closed form, ${{\bf{dF}}=0\,}$. Note that since the action Eq.~\eqref{action3F} bears a self-interaction term, $V(A^2)$, gauge invariance may be broken. However it is possible to restore this symmetry through the introduction of a St\"{u}ckelberg form \cite{Koivisto:2009sd,Koivisto:2009fb}.

The total action of our theory minimally coupled to Einstein's gravity, can thus be written as
\begin{equation}
\label{action}
\mathcal{S} = \int \omega_g \left[ \frac{1}{16\pi G}R + \mathcal{L}_m (g_{\mu\nu},\psi,\nabla\psi) \right]+\mathcal{S}_A,
\end{equation}
where $G$ is Newton's constant, $R$ the curvature scalar and $\mathcal{L}_m$ stands for an
anisotropic distribution of matter threading our spactime. The field action $\mathcal{S}_A$ is given
by ${\mathcal{S}_A = \int \omega_g\,\mathcal{L}_A}$, where $\mathcal{L}_A$ is the field's
Lagrangian density.

To find the dynamics governing this model, we start by varying the action Eq.~\eqref{action} with respect to the three-form, and find the following equations of motion:
\begin{equation}
\label{motionT}
E_{\alpha\beta\gamma} = \nabla_{\mu}F^{\mu}{}_{\alpha\beta\gamma} -12\frac{\partial V}{\partial (A^2)}A_{\alpha\beta\gamma}=0.
\end{equation}
The energy-momentum tensor relative to the three-form reads
\begin{eqnarray}
T^{(A)}_{\mu\nu} &=& -2 \frac{\delta \mathcal{L}_{A}}{\delta g^{\mu\nu}}+g_{\mu\nu}\mathcal{L}_{A} \nonumber \\
&=& \frac{1}{6}F_{\mu\alpha\beta\gamma}F_{\nu}{}^{\alpha\beta\gamma} + 6 \frac{\partial V}{\partial (A^2)} A_{\mu\alpha\beta}A_{\nu}{}^{\alpha\beta} +\mathcal{L}_A\,g_{\mu\nu},\quad \label{emtensor}
\end{eqnarray}
Note, however, that the equations of motion \eqref{motionT} could equivalently be deduced from the contracted Bianchi identities \cite{Wongjun:2016tva}
\begin{equation}
\nabla_{\mu}T^{(A)\,\mu}{}_{\nu}=\frac{1}{6}F_{\nu\alpha\beta\gamma}E^{\alpha\beta\gamma}=0.
\end{equation}

The modified field equations can be computed by varying Eq.~\eqref{action} with respect to $g^{\mu\nu}$, leading to
\begin{equation}
\label{field}
G_{\mu\nu} = 8\pi G\left( T^{(m)}_{\mu\nu} + T^{(A)}_{\mu\nu} \right),
\end{equation}
with $G_{\mu\nu}$ being the components of the Einstein tensor.

Taking the divergence of Eq. \eqref{field} and using the field equation %
\eqref{motionT} one can find that the energy-momentum tensor of matter is conserved
\begin{align}  \label{cons}
\nabla^\mu\, T^{(m)}_{\mu\nu}=0.
\end{align}
Regarding the matter source, we assume an anisotropic distribution of
matter, where the components of the energy-momentum tensor can be written
as,
\begin{equation}
T^{(m)}_{\mu\nu} = (\rho_m+p_{\perp m})u_\mu u_{\nu} + p_{\perp m} g_{\mu\nu} +
(p_{rm}-p_{\perp m})\chi_{\mu}\chi_{\nu},
\end{equation}
$u_{\mu}$ being the four-velocity, normalized as $u_{\mu}u^{\mu}=-1$, $%
\chi^{\mu}$ the unit spacelike vector in the radial direction, $\rho_m$, $p_{rm}$ and $p_{\perp m}$ the energy density, radial pressure and tangential pressure, respectively.

Note that this present theory considers standard General Relativity for the gravitational Lagrangian with the novelty being the introduction of a 3-form field fluid in the matter sector which modifies the overall dynamics through the intimate relation of the gravitational field equations \eqref{field}. Thus, we will informally refer to this model as three-form gravity.
%

\section{Spherically symmetric and static background}\label{back}

This work aims at finding solutions for three-form supported stars. To this effect, let us consider the following static and spherically symmetric line element \cite{Lobo:2005uf}
\begin{eqnarray}
ds^2 = &-&\exp\left[-2\int_r^{\infty}h(\tilde{r})d\tilde{r}\right]dt^2 +
\frac{dr^2}{1-2\, G \,m(r)/r}  \nonumber \\
&+& r^2 \left( d\theta^2 + \sin^2\theta \,d\phi^2 \right),  \label{metric}
\end{eqnarray}
where $m(r)$ is the mass function, and $h(r)$ is the gravity profile, both
functions of the radial coordinate, $r$, only.

Similar to previous studies \cite{Barros:2018lca,Barros:2020ghz} on static and spherically symmetric backgrounds, we will assume an ansatz for our three-form field through the aid of its Hodge dual 1-form ${{\bf B}}=\star{{\bf A}}$, which fully characterize the components of the field, i.e.,
\begin{equation}
\label{dual}
A_{\alpha\beta\gamma} = \sqrt{-g}\,\epsilon_{\alpha\beta\gamma\delta}B^{\delta}.
\end{equation}
Hence, we choose a convenient parametrization of ${{\bf B}}$ as the radially directed vector with components,
\begin{equation}
B^{\delta}=\left( 0,\sqrt{1-\frac{2Gm(r)}{r}}\zeta(r),0,0 \right)\,.
\end{equation}
dependent on a suitable scalar function $\zeta (r)\,$. Accordingly, the dynamical equations for the three-form will be expressed entirely in terms of $\zeta(r)$. Since this scalar function fully determines the components of the three-form, by way of Eq.~\eqref{dual}, throughout this manuscript we may naively refer to $\zeta$ simply as the three-form, although, formally, it is the scalar function determining the components of the three-form, $A_{\alpha\beta\gamma}$. We refer the reader to appendix A of \citep{DeFelice:2012jt} to examine how the action Eq.~\eqref{action3F} may be written solely in terms of the dual vector ${{B}}$, highlighting how this model can be recast into a vector-tensor theory at the background level.

In this gravitational setting, the contraction $A^2$ becomes:
\begin{equation}\label{aux1}
A^2 = -6B^{\delta}B_{\delta} = -6 \,\zeta (r)^2,
\end{equation}
and the kinetic invariant
\begin{eqnarray}
\label{F2}
F^2  &:=& F^{\alpha\beta\gamma\delta}F_{\alpha\beta\gamma\delta} = -24\,(\nabla_{\delta}B^{\delta})^2 \nonumber \\
&=& -24\left( 1-\frac{2Gm}{r} \right)\left[ \zeta\left( \frac{2}{r} - h \right) + \zeta' \right]^2,
\end{eqnarray}
where derivatives with respect to the radial coordinate $r$ are denoted by a prime and $\zeta = \zeta(r)$. We may now express the dynamics of the three-form, Eq.~\eqref{motionT}, using the metric Eq.~\eqref{metric}, solely in terms of $\zeta$ and the metric functions, through:
\begin{gather}  \label{motion}
r^2\zeta^{\prime \prime }\left( \frac{2Gm}{r}-1 \right) +r\,\zeta^{\prime }%
\left[ G m^{\prime }+\frac{3Gm}{r}-2 \right. \notag\\
\left.+rh\left( 1-\frac{2Gm}{r} \right) \right]+\zeta\left[ 2+2G \left(m^{\prime }- \frac{3m}{r}\right)\right. \notag\\
\left.+rhG\left( \frac{m}{r}-m^{\prime
}\right)+r^2 h^{\prime }\left( 1-\frac{2Gm}{r} \right) \right]-r^2V_{,\zeta}=0,
\end{gather}
with $V_{,\zeta} = \partial V / \partial \zeta\,$.

The components of the energy-momentum tensor of the three-form can be computed by plugging Eq.~\eqref{metric} in Eq.~\eqref{emtensor}, which provides:
\begin{eqnarray}
T^{(A)\,\,t}{}_{t} &=& -\rho_A = \frac{1}{48}F^2 - V + \zeta
V_{,\zeta}, \\
T^{(A)\,\,r}{}_{r} &=& p_{rA} = \frac{1}{48}F^2 - V, \\
T^{(A)\,\,\theta}{}_{\theta} &=& T^{(A)\,\,\phi}{}_{\phi}
= p_{\perp A} = T^{(A)\,\,t}{}_{t},  \label{T3}
\end{eqnarray}
with the object $F^2$ given by Eq.~\eqref{F2}.

The components of the modified field equations, Eqs.~\eqref{field}, under such a gravitational framework yield
\begin{eqnarray}
\rho_A + \rho_m &=& \frac{m^{\prime }}{4\pi r^2},  \label{fieldI} \\
p_{rA}+p_{rm} &=&- \frac{m}{4\pi r^3}-\frac{h}{4\pi G r}\left(1- \frac{2G m}{r}\right),
\label{fieldII} \\
p_{\perp A}+p_{\perp m} &=& \frac{1}{8\pi G}\left( 1-\frac{2G m}{r} \right)\left[\left( h-\frac{1}{r} \right) \times \right.  \notag \\
&&\left. \times\left( h+\frac{Gm'r-Gm}{r(r-2Gm)} \right) -h'\right].\label{fieldIII}
\end{eqnarray}

In addition to this, the conservation equation \eqref{cons} takes the form
\begin{align}  \label{cons1}
p_{rm}^{\prime }=\frac2r\left(p_{\perp
	m}-p_{rm}\right)+h\left(\rho_m+p_{rm}\right).
\end{align}

By eliminating $h$ between Eqs.~(\ref{fieldII}) and (\ref{cons1}) we obtain
the generalized Tolman-Oppenheimer-Volkoff (TOV) equation describing the
structure of massive compact objects in three-form gravity, given by
\begin{eqnarray}  \label{TOV}
p^{\prime }_{rm}&=&-\frac{G\left(\rho _m+p_{rm}\right)\left[%
	\left(p_{rA}+p_{rm}\right)4\pi r^3+m\right]}{r^2\left(1-2Gm/r\right)}  \notag \\
&&+\frac{2}{r}\left(p_{\perp m}-p_{rm}\right).
\end{eqnarray}

The TOV equation, together with the mass continuity equation (\ref{fieldI}),
and the equation of motion for $\zeta$, form a system of differential
nonlinear ordinary equations, whose solution fully characterize star-like
objects. The system must be considered with the initial conditions $m(0)=0$,
$p_{rm}(0)=p_{rm}^{(0)}$, and $\zeta (0)=\zeta _0$, respectively. In order to
close the system of equations the equations of state of the dense matter and
the functional form of the potential $V\left(A^2\right)$ must also be
specified.

We want to express the results in units of $km$ and $M_{\odot}$, and hence we use a set of
dimensionless quantities defined as
\begin{align}
&\rho_m= \epsilon_0 \bar{\rho}_m, \quad p_{rm}= \epsilon_0 \bar{p}_{rm},
\quad p_{\perp m}= \epsilon_0 \bar{p}_{\perp m},~~ h=\bar{h}/R_0, \notag\\ & m=M_\odot\,\bar{m}, \quad\zeta=\bar{\zeta}\left(M_\odot/R_0\right)^{1/2},\quad r= R_0 \,\eta,
\end{align}
where $\epsilon_0$ is an arbitrary energy density scale, $M_\odot$ is solar mass, and $%
R_0=G M_\odot/c^2=1.477 ~{\rm km}$. In the following we set $\epsilon_0=M_\odot c^2/80 \pi R_0^3=2.45\times10^{15} \, {\rm g/cm}^3$. One should note that according to the above definitions  we have
\begin{align}
&\rho_A= \epsilon_0 \bar{\rho}_A, \qquad p_{rA}= \epsilon_0 \bar{p}_{rA},
\qquad p_{\perp A}= \epsilon_0 \bar{p}_{\perp A}.
\end{align}

In the dimensionless variables defined above the system of the structure
equations of the three-form stars take the form
\begin{equation}\label{mcont}
\frac{d\bar{m}}{d\eta }=\frac{1}{20}\left( \bar{\rho}_{A}+\bar{\rho}_{m}\right) \eta
^{2},
\end{equation}%
\begin{eqnarray}\label{TOVdim}
\frac{d\bar{p}_{rm}}{d\eta } &=&-\frac{\left( \bar{\rho}_{m}+\bar{p}%
	_{rm}\right) \left[ \left( \bar{p}_{rA}+\bar{p}_{rm}\right) \eta ^{3}/2+ 20 \,\bar{%
		m}\right] }{20\, \eta ^{2}\left( 1-2\bar{m}/\eta \right) }  \notag \\
&&+\frac{2}{\eta }\left( \bar{p}_{\perp m}-\bar{p}_{rm}\right) ,
\end{eqnarray}%
\begin{eqnarray}%
&\eta ^{2}\frac{d^{2}\bar{\zeta} }{d\eta ^{2}}\left( \frac{2\bar{m}}{\eta }%
-1\right) +\eta \frac{d\bar{\zeta} }{d\eta }\,\left[ \frac{d\bar{m}}{d\eta }+\frac{%
	3\bar{m}}{\eta }-2+\eta \bar{h}\left( 1-\frac{2\bar{m}}{\eta }\right) \right]
\notag \\
&+\bar{\zeta} \left[ 2+2\frac{d\bar{m}}{d\eta }-\frac{6\bar{m}}{\eta }+\eta \bar{h}%
\left( \frac{\bar{m}}{\eta }-\frac{d\bar{m}}{d\eta }\right) +\eta ^{2}\frac{d%
	\bar{h}}{d\eta }\left( 1-\frac{2\bar{m}}{\eta }\right) \right]   \notag \\
&-\frac{1}{80 \pi}\frac{d\bar{V}\left( \bar{\zeta}\right) }{d\bar{\zeta}} \eta ^{2}=0.\label{zetadim}
\end{eqnarray}%

The system of equations (\ref{mcont})--(\ref{zetadim}) must be integrated with the boundary conditions $\bar{m}(0)=0$, $\bar{p}_{rm}(0)=p_{rm}(0)/\epsilon _0$, $\bar{p}_{\perp m}(0)=p_{\perp m}(0)/\epsilon _0$ etc., $\bar{\zeta}(0)=\bar{\zeta} _0$, $\bar{\zeta}'(0)=\bar{\zeta} ^{\prime}_0$, and $\bar{p}_{rm}(R)=0$, respectively, with the latter condition determining the radius of the compact stellar objet in the three-form fields gravity theory.

In the following we are going to construct specific models of three-form field stars under the assumption that the three form field vanishes at the center of the star, and increases towards the surface, where it reaches its maximum value. Therefore, we impose the condition $\zeta (0)=0$, indicating that the field becomes negligibly small near the center. Hence, from a physical point of view we assume that the three-form field, as well as the associated physical parameters, do follow, at least qualitatively, the variation of the mass distribution inside the star, and that the maximum values of the field are related to the maximum values of the mass of the ordinary matter. Moreover, we assume a slow increase of the field near the center of the star, with $\zeta '(0)\neq 0$, with the field derivative also taking a small value near the center $r=0$, so that $\zeta '(0) \ll 1$. Indeed, different types of models in which the field is, for example, maximum at the center of the star, or it has some arbitrary large values at $r=0$,  can also be constructed.

\section{Structure and astrophysical properties of compact objects}\label{struct}

In the present Section, we will investigate the basic astrophysical properties of high density compact objects  in the three-form field gravity. In our analysis we will not impose any specific restrictions on the functional form of the three-form field $\zeta$, which is assumed to obey Eq.~ (\ref{zetadim}). In order to close the system of field equations describing the interior of the stellar objects we need to adopt an equation of state for the ordinary matter. Despite intensive theoretical efforts the equation of state of high density baryonic matter is poorly known, and a realistic description of the physical properties of matter at densities exceeding with several orders of magnitude the nuclear density $\rho _n=2\times 10^{14}$ g/cm$^3$ is still missing. For this reason we consider some simple approximations of the matter equation of state, by considering four specific cases, given by: (i) the stiff fluid equation of state, in which the pressure equals the energy density, $P=\rho$; (ii) the radiation fluid equation of state corresponding to $P = \rho/3$; (iii) the important quark matter equation of state $P= \left(\rho-4B\right) /3$, where $B$ is the bag constant,  and (iv) the superfluid neutron matter Bose-Einstein Condensate equation of state, which is given by a polytropic equation of state with the polytropic index $n=1$, so that $P\propto \rho ^2$.

An important physical quantity with a profound influence on the stellar structure is the three-form potential $V\left(A^2\right)$. In the following we will restrict our investigations to two functional forms of $V\left(A^2\right)$. We will assume that it is either a constant, or it has a Higgs-type form, with $V\left(A^2\right)=aA^2+bA^4$, respectively, where $a$ and $b$ are constants. From a physical point of view the constant $a<0$ is related to
the mass of the three-form field via the relation $m_{\zeta}^2= bv^2/2=-2a$, where $v^2=-a/2b$ gives the minimum of the Higgs-type potential.

\subsection{Stellar structures with stiff matter fluid}

As a first example of a stellar model in the three-form field gravity we consider the case of an isotropic distribution of matter, with $p_{rm}=p_{\perp m}$, with the matter pressure satisfying the stiff fluid (Zeldovich) equation of state \cite{Zeldovich:1962emp,Shapiro:1983du}
\begin{align}
p_{rm}=\rho_m.
\end{align}

From a physical point of view such an equation of state may describe the matter behavior at densities ten times higher than the nuclear density, i.e., at densities greater than or equal to $10^{17}$ g/cm$^3$, corresponding to temperatures $T=(\rho/\sigma)^{
1/4} >10^{13}$ K, where by $\sigma$ we have denoted the radiation constant \cite{Shapiro:1983du}. An important characteristic of the Zeldovich equation of state is that for stiff matter the speed of sound is equal to the speed of light, $c_s^2=\partial p_{rm}/\partial \rho _m=1$. One reason why such an equation of state was proposed is that for stiff matter the matter perturbations cannot move at speeds greater than the speed of light. The stiff matter equation has found many applications in astrophysics. In \cite{Rhoades:1974fn} it was shown that for the equilibrium configuration of
a stellar object with a very high density the maximum mass cannot surpass the upper limit value of $3.2M_{\odot}$. To obtain this fundamental result it was assumed that the stellar interior is described by the spherically symmetric static Einstein field equations, and that the principle of
causality, and Le Chatelier’s principle both hold. Moreover, it was assumed
that at very high densities the neutron matter satisfies the stiff fluid equation of state $p=\rho$. This important result on the numerical value of the maximum mass of stable compact objects offers a fundamental criterion for observationally distinguishing ordinary neutron or other types of compact stars from black holes, and it represents an important prediction of the theory of general relativity.

In considering the properties of three-form field stars described by the stiff matter equation of state  we assume that the central density of the compact object varies between the values $3.1\times10^{14}\;{\rm  g/cm}^3$ and $2.2\times 10^{15}\; {\rm g/cm}^3$. We stop the numerical integration when the density reaches the surface value $\rho_{m}= 10^{14}\; {\rm g/cm}^3$, a density lower than the nuclear density. The initial value of the three-form field and of its derivative is $\bar{\zeta}_0=0$ and $\bar{\zeta}^\prime_0=7\times 10^{-3}$
in all considered cases. In the following, we present the results of the numerical integration for stiff fluid stars obtained by considering two different
choices of the three-form field potential to obtain the interior solutions.

\subsubsection{The constant potential case: $V(A^2)=\lambda$}

As a first case of a stellar type structure constructed in the three-form field gravity we assume that the field potential is a constant, $V(A^2)=\lambda={\rm constant}$. To describe the properties of the potential we introduce the dimensionless parameter $\bar{\lambda}$, defined as
\begin{align*}
\lambda=\epsilon_0 \bar{\lambda}.
\end{align*}

\begin{figure*}[htbp]
\centering
\includegraphics[width=8.3cm]{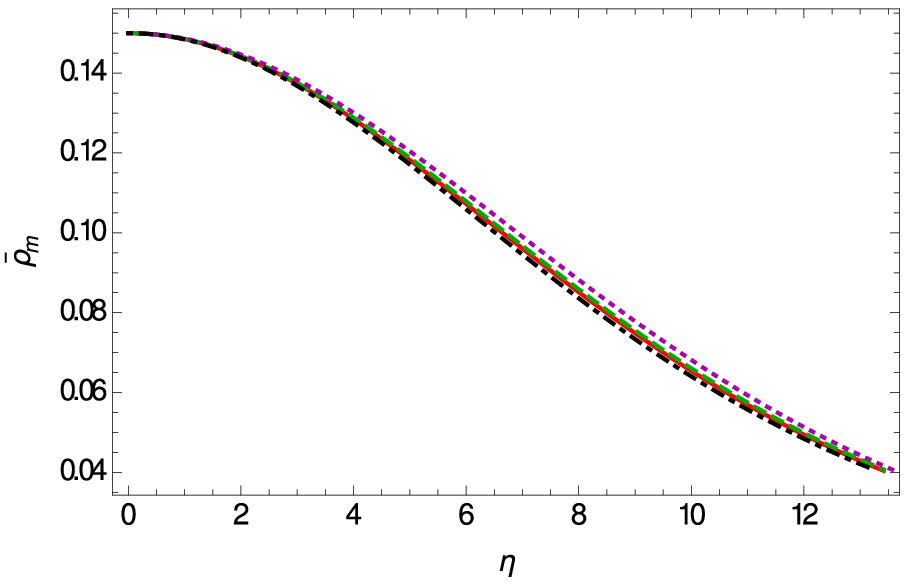}\hspace{.4cm}
\includegraphics[width=8.0cm]{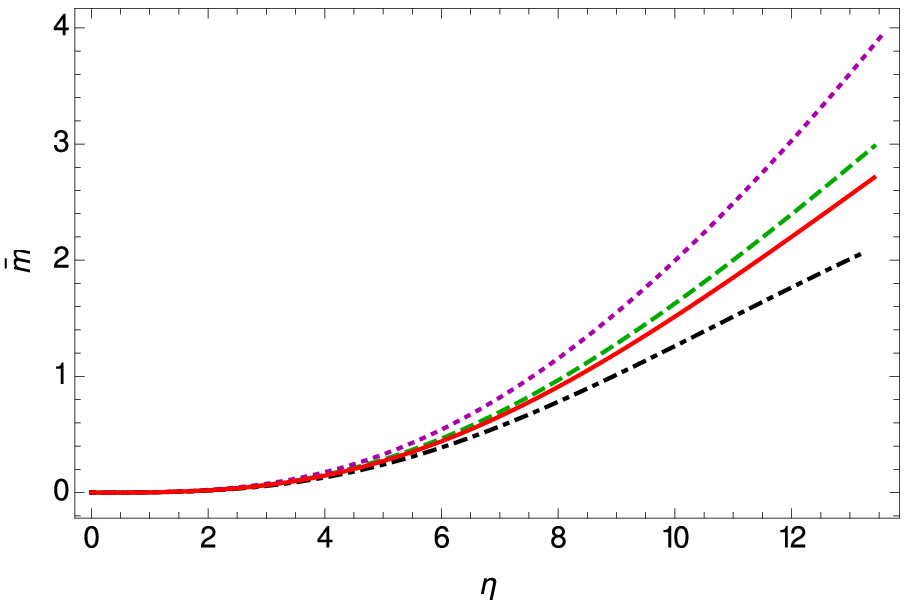}
\caption{The interior mass energy density profile $\bar{\rho}_m$ (left panel) and the mass profile $\bar{m}$ (right figure) as a function of the distance from the center of the
three-form  stiff fluid stars $\eta$, in the presence of a constant field potential, for three different values of the constant $\bar{%
\lambda}$:  $\bar{\lambda}=0$ (dashed curve),  $\bar{\lambda}=0.02$ (dotted curve), and  $\bar{%
\lambda}=-0.02$ (dot-dashed curve), respectively.  For the central density of the star we have adopted the value $\rho_{mc}=3.7\times 10^{14} {\rm g/cm}^3$. The solid curve represents the standard general relativistic density and mass profiles for stiff fluid stars. }
\label{fig1}
\end{figure*}

The variations with respect to the dimensionless coordinate $\eta$ of the energy density of the matter and of the mass of the three-form stars are represented, for different values of $\bar{\lambda}$, in Fig.~\ref{fig1}.
As one can see from the plots, the matter energy density is a monotonically decreasing function of the radial coordinate, and it tends towards  the zero value at the star surface. The dependence on the numerical values of the potential $\bar{\lambda}$ is weak, at least for small values of the radial coordinate. The interior mass profile of the star is a monotonically increasing function of $\eta$, and it shows a significant dependence on $\bar{\lambda}$.

\begin{figure*}[htbp]
\centering
\includegraphics[width=8.3cm]{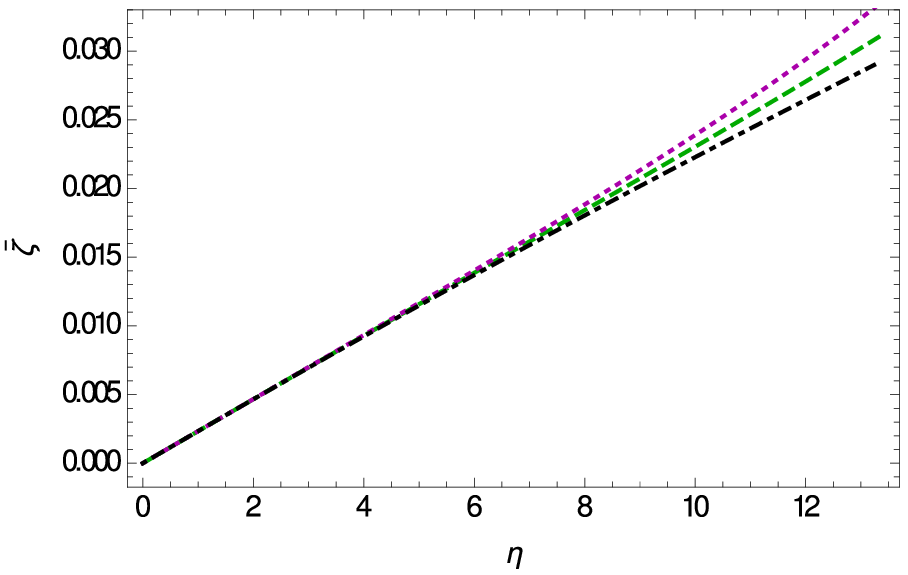}\hspace{.4cm}
\includegraphics[width=8.0cm]{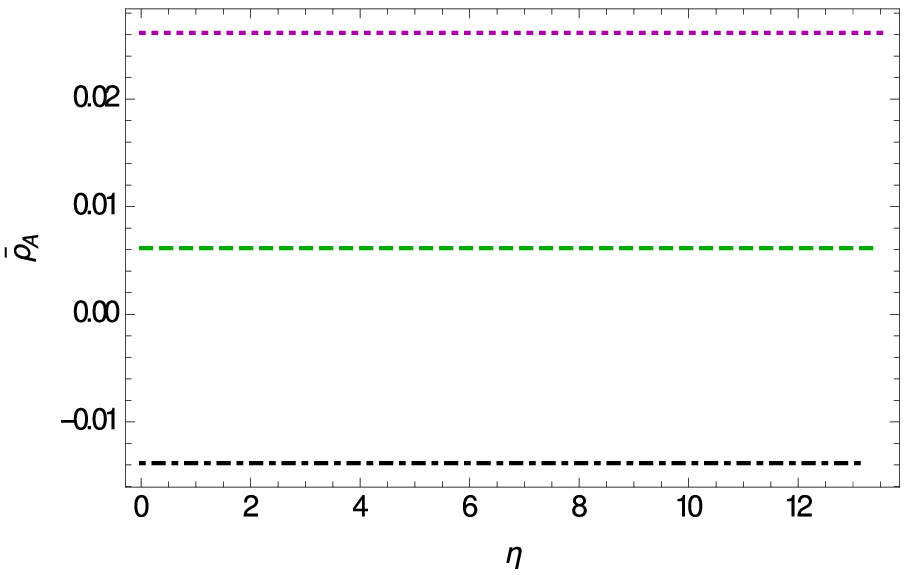}
\caption{Variation of the non-zero component  $\bar{\zeta}$ (left panel) and of the effective energy density  $\bar{\rho}_A$ (right figure) of the three-form field as a function of the radial
distance from the center of the stiff fluid star $\eta$ with constant potential, for three different values of the
constant $\bar{\lambda}$: $\bar{\lambda}=0$ (dashed curve),  $\bar{\lambda}=0.02$ (dotted curve), and  $\bar{%
\lambda}=-0.02$ (dot-dashed curve), respectively.  For the central density of the star we have adopted the value $\rho_{mc}=3.7\times 10^{14} {\rm g/cm}^3$, while $\bar{\zeta}_0=0$ and $\bar{\zeta}^\prime_0=7\times 10^{-3}$, respectively. }
\label{fig2}
\end{figure*}

The variations  of the three-form field $\bar{\zeta}$ and of the energy density of the three-form fields for the stiff fluid stars in the presence of a constant potential are depicted in Fig.~\ref{fig2}. The three-form field $\bar{\zeta}$ is a monotonically increasing function inside the star, and it reaches its maximum value on the star surface. On the other hand, inside the star the energy density of the three-form field is a constant, independently of the adopted values of the parameters. Moreover, depending on the numerical values of the three-form field potential,  $\bar{\rho}_A$ can take both negative and positive values.

\begin{figure}[hh!]
\centering
\includegraphics[width=7.5cm]{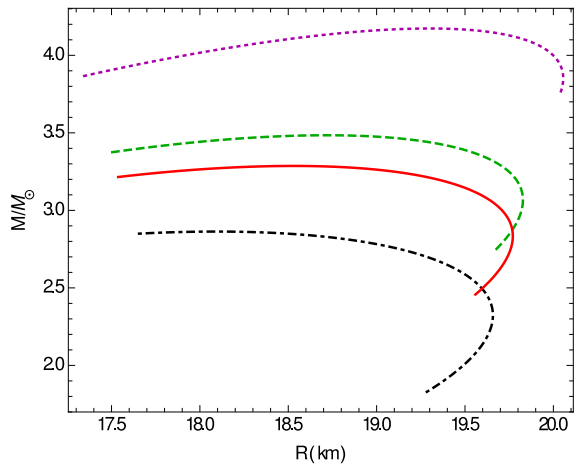}
\caption{Variation of $M/M_{\odot}$ as a function of the radius of the star $R$ (km)
for three-form stiff fluid stars, for three different values of the constant $\bar{%
\lambda}$ for three different values of the constant $\bar{%
\lambda}$:  $\bar{\lambda}=0$ (dashed curve),  $\bar{\lambda}=0.02$ (dotted curve), and  $\bar{%
\lambda}=-0.02$ (dot-dashed curve), respectively. The solid curve represents the mass-radius relation for general relativistic stiff fluid stars. }
\label{fig3}
\end{figure}

The mass-radius relations of the three-form stiff fluid stars with constant potential are shown in Fig.~\ref{fig3}. The general relativistic case is also represented for comparison. As one can see from the Figure, the mass-radius relation for compact objects shows significant differences as compared to the standard general relativistic case. Much higher maximum masses, of the order of $4.1M_{\odot}$, exceeding the maximum mass limit of $3.2M_{\odot}$ of general relativity, can be achieved for positive values of the three-form field potential. On the other hand negative values of $\bar{\lambda}$ lead to a decrease of the maximum mass of the neutron stars to a value of the order of $2.5M_{\odot}$, and generally of the values of the stellar masses.

Some specific numerical values of the maximum masses of stiff fluid stars in three-form field gravity are presented, for different values of the field potential, in Table~\ref{stif-vcons-tab}. In general relativity for stiff fluid stars we have ${M_{max}=3.28\,M_{\odot}}$, $R=18.53\,{\rm km} $, and ${\rho_{mc}=1.21\times 10^{15}\, {\rm g/cm}^3}$, respectively.

\begin{table}[hh!]
	\begin{center}
		\begin{tabular}{|c|c|c|c|}
			\hline
			$\bar{\lambda}$ &~~~$-0.02$~~~&$~~~0.0~~~~$&$~~~0.02~~~~$ \\
			\hline
			$~\rho_{mc} \times 10^{-14}\,({\rm g/cm}^3)~$& $~~~16.6~~~$& $~~~10.6~~~$& $~~~6.89~~~$\\
			\hline
			\quad$M_{max}/M_{\odot}$\quad& $~~~2.86~~~$& $~~~3.48~~~$& $~~~4.17~~~$\\
			\hline
			$~~~R\,({\rm km})~~~$& $~~~18.09~~~$& $~~~18.71~~~$& $~~~19.30~~~$\\
			\hline
		\end{tabular}
		\caption{The maximum mass and corresponding radius for three-form field stiff-fluid stars with constant potential.}\label{stif-vcons-tab}
	\end{center}
\end{table}

\subsubsection{Higgs-type potential: $V(A^2)=a A^2 +b A^4$}

For the case of stiff fluid stars with a Higgs-type three-form field potential given by $V(A^2)=a A^2 +b A^4$, we rescale the potential parameters to a dimensionless form so that
\begin{align*}
\bar{a}=R_0^2\, a, \quad \bar{b}= R_0 M_\odot b.
\end{align*}

\begin{figure*}[htbp]
\centering
\includegraphics[width=8.3cm]{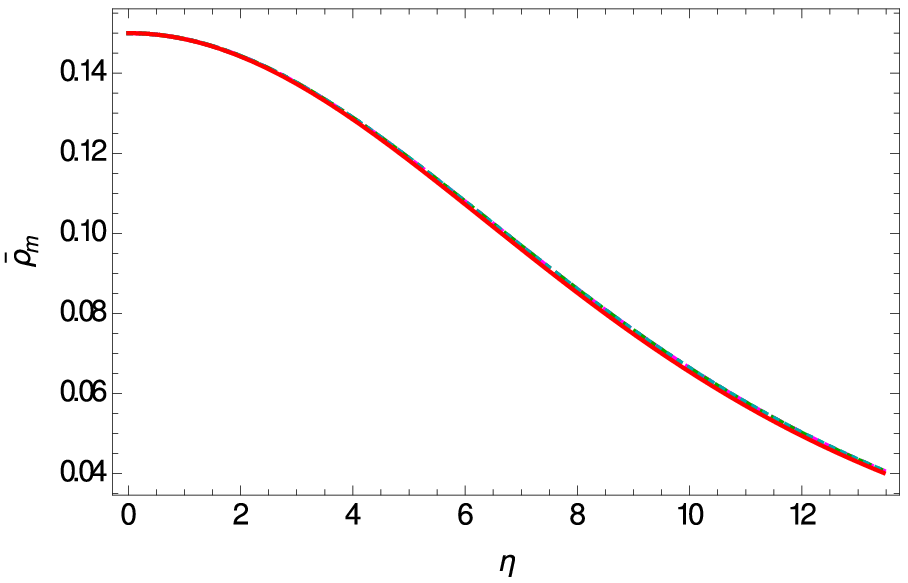}\hspace{.4cm}
\includegraphics[width=8.0cm]{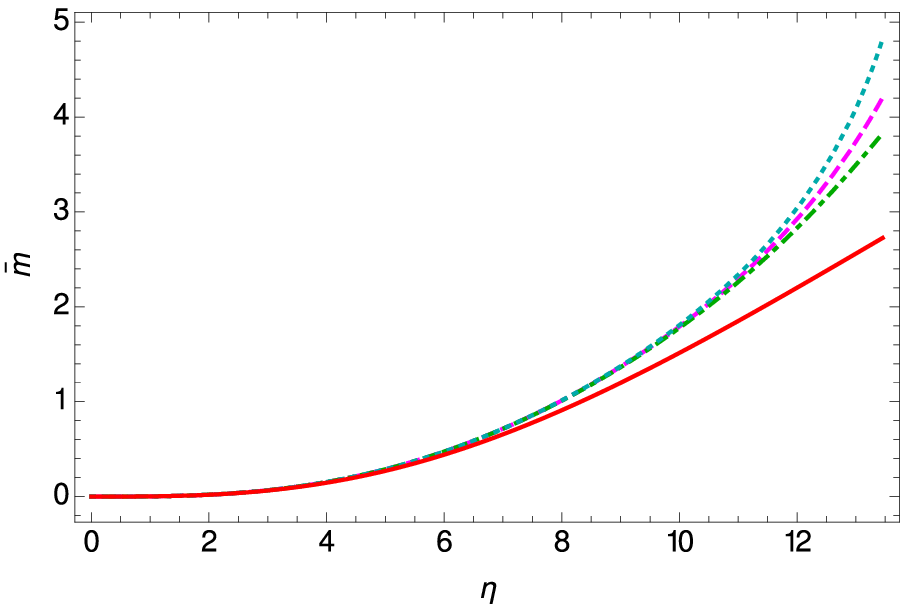}
\caption{Variations of $\bar{\rho}_m$ (left panel), and of $\bar{m}$ (right panel) as a function of $\eta$ for three-form stiff fluid stars in the presence of a Higgs type potential, for $\bar{a}%
=0.003$ and for three different values of the constant $\bar{b}$:  $\bar{b}=0$ (dashed curve),  $\bar{b}=-0.033$ (dotted curve), and
 $\bar{b}=0.033$ (dot-dashed curve). For the central density of the star we adopt the value $\rho_{mc}=3.7\times 10^{14} {\rm g/cm}^3$. The solid curve represents the standard general relativistic density and mass profiles for stiff fluid stars. }
\label{fig4}
\end{figure*}

In the following, we consider the cases where $\bar{a}=0.004$, and $\bar{b}=0,\pm 0.01$, respectively. The variations of the interior density and mass profiles are represented in Fig.~\ref{fig4}. As required by physical considerations, the energy density of the stiff fluid three-form star in the presence of the Higgs potential is a monotonically decreasing function of the radial coordinate. The variation of the energy density is practically independent on the values of the parameters of the Higgs potential and, at least for the considered range of values, it is very similar to the general relativistic case. However, there is a significant effect on the variation of the mass profile of the potential parameters, with higher mass values obtained for higher values of $\bar{b}$.

\begin{figure*}[htbp]
\centering
\includegraphics[width=7.8cm]{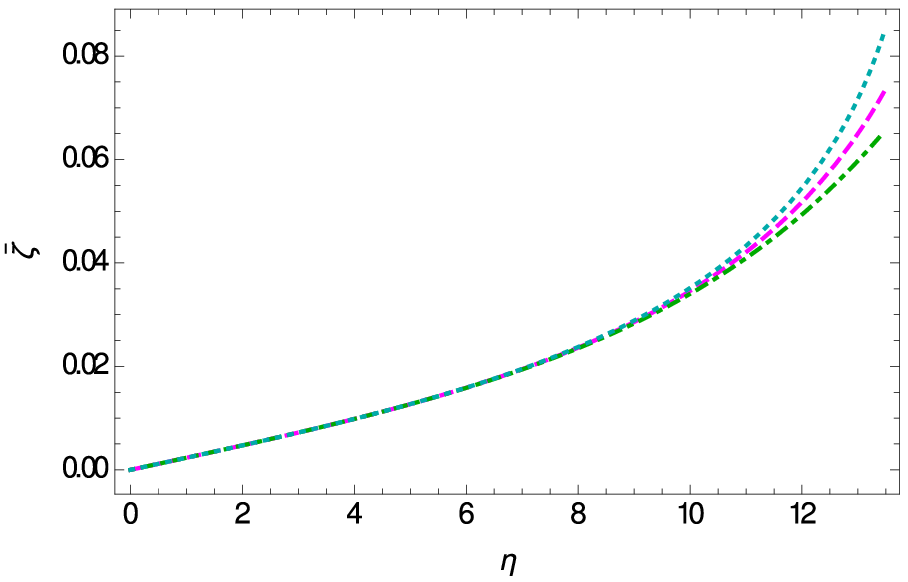}\hspace{.4cm}
\includegraphics[width=8.0cm]{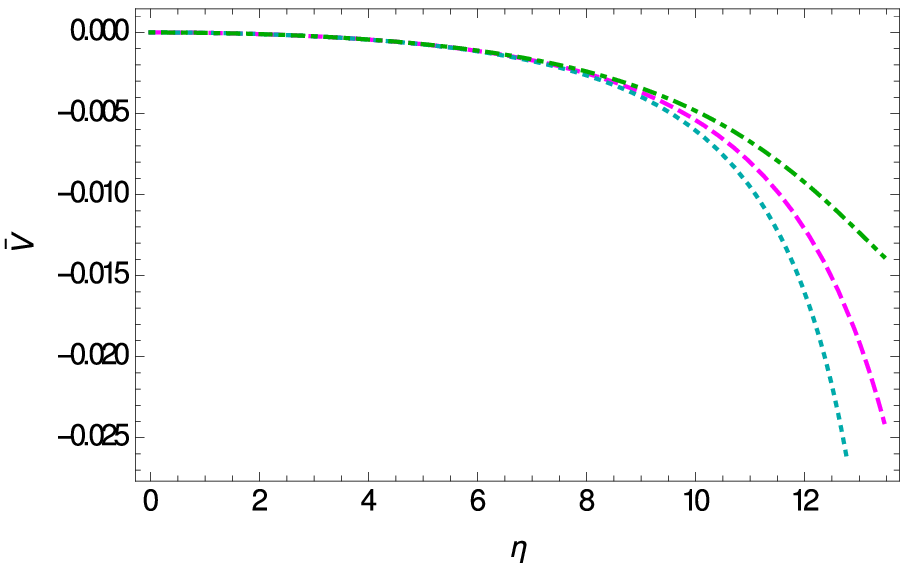}
\caption{Variations of $\bar{\zeta}$ (left panel) and of $\bar{V}\left(\bar{\zeta}\right)$ (right panel) as functions of $\eta$ for stiff fluid three-form stars in the presence of the Higgs potential, for $\bar{a}%
=0.003$ and three different values of the Higgs potential parameter $\bar{b}$: $\bar{b}=0$ (dashed curve),  $\bar{b}=-0.033$ (dotted curve), and
 $\bar{b}=0.033$ (dot-dashed curve). For the central density of the star we adopt the value $\rho_{mc}=3.7\times 10^{14} {\rm g/cm}^3$, while $\bar{\zeta}_0=0$ and $\bar{\zeta}^\prime_0=7\times 10^{-3}$, respectively. }
\label{fig5}
\end{figure*}

\begin{figure}[htbp]
	\centering
	\includegraphics[width=7.8cm]{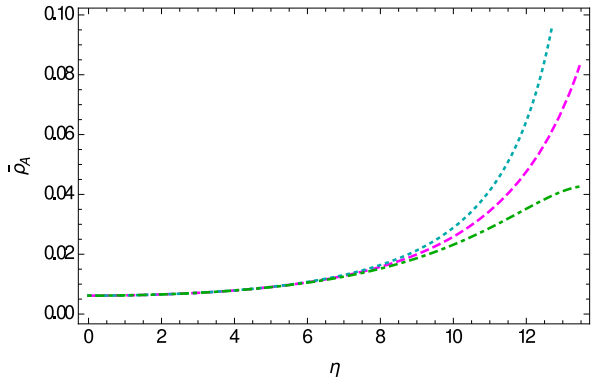}
	\caption{Variations of $\bar{\rho}_A$  as function of $\eta$ for stiff fluid three-form stars in the presence of the Higgs potential, for $\bar{a}%
		=0.003$ and three different values of the Higgs potential parameter $\bar{b}$: $\bar{b}=0$ (dashed curve),  $\bar{b}=-0.033$ (dotted curve), and
		$\bar{b}=0.033$ (dot-dashed curve). For the central density of the star we adopt the value $\rho_{mc}=3.7\times 10^{14} {\rm g/cm}^3$, while $\bar{\zeta}_0=0$ and $\bar{\zeta}^\prime_0=7\times 10^{-3}$, respectively. }
	\label{densa-stif-vvar}
\end{figure}

Similarly to the constant potential case, the three-form field component is a monotonically increasing function inside the star, reaching its maximum value near the star surface (see Fig.~\ref{fig5} for details). There is a significant dependence of $\bar{\zeta}$  on the parameters of the Higgs potential near the vacuum boundary. The Higgs-type potential is a monotonically decreasing function of $\eta=r/R_0$, as is transparent from Fig.~ \ref{fig5} and, at least for the considered set of parameters, it has only negative values, reaching its minimum on the star surface. The variation of the effective energy density of the three-form field is represented in Fig.~\ref{densa-stif-vvar}. As one can see from the Figure, $\bar{\rho}_A$ is a monotonically increasing function of $\eta $ inside the star, reaching its maximal value at the vacuum boundary.

\begin{figure}[htbp]
\centering
\includegraphics[width=7.8cm]{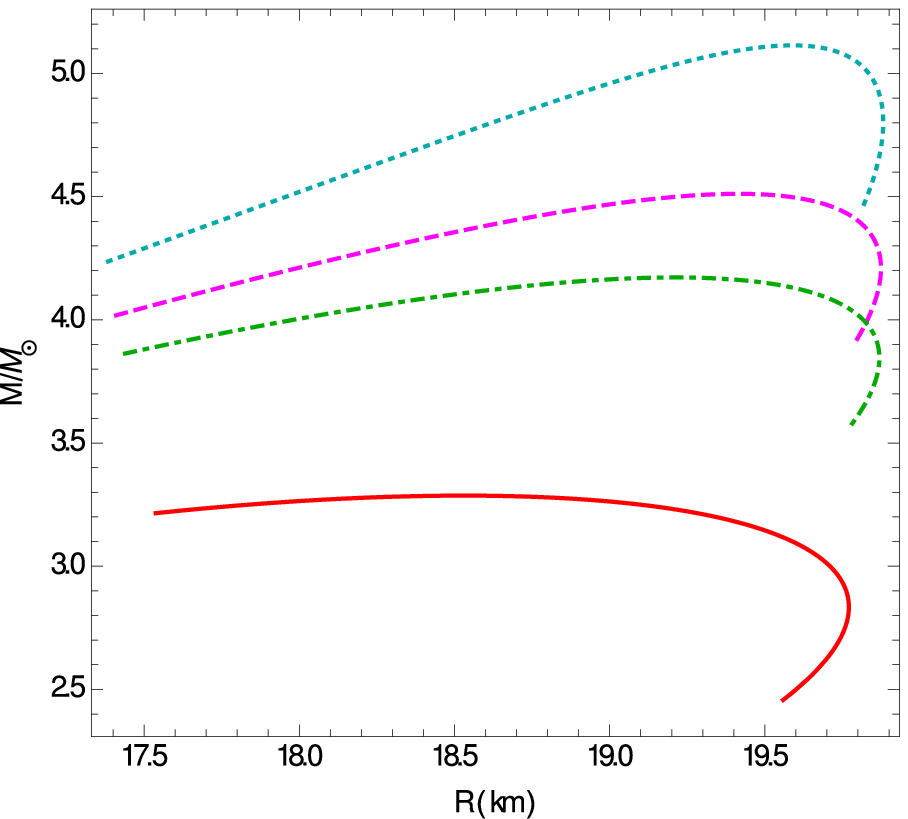}
\caption{Variation of the mass $M/M_{\odot}$ as a function of the radius of the star $R$ for a three-form stiff fluid star in the presence of a Higgs type potential,
for $\bar{a}=0.003$ and three different values of the constant $\bar{b}$: $\bar{b}=0$ (dashed curve),  $\bar{b}=-0.033$ (dotted curve), and
 $\bar{b}=0.033$ (dot-dashed curve). The solid curve represents the mass-radius relation for general relativistic stiff fluid stars. }
\label{fig6}
\end{figure}

The mass-radius relation for three-form field stiff fluid stars in the presence of a Higgs-type potential is shown in Fig.~\ref{fig6}.
As one can see from Fig.~\ref{fig6}, three-form field compact objects obeying the stiff fluid equation of state can have much higher masses than their general relativistic counterparts, with the maximum mass reaching values as high as $5.10M_{\odot}$. Generally, for all considered numerical values the stiff fluid three-form stars in the presence of a Higgs potential have higher masses as compared to the standard general relativistic approach. The general shape of the mass-radius dependence curve is similar to the one in general relativity, with a displacement of the curve towards higher mass values.

Exact numerical values of the maximum mass and of the corresponding radius for stiff fluid stars in the presence of a Higgs-type potential are given in Table~\ref{stif-vvar-tab}.

\begin{table}[h!]
	\begin{center}
		\begin{tabular}{|c|c|c|c|}
			\hline
			$\bar{b}$ &~~~$-0.033$~~~&$~~~0.0~~~~$&$~~~0.033~~~~$ \\
			\hline
			$~~~\rho_{mc} \times 10^{-14}\,({\rm g/cm}^3)~~~$& $~~~5.50~~~$& $~~~6.38~~~$& $~~~7.45~~~$\\
			\hline
			\quad$M_{max}/M_{\odot}$\quad& $~~~5.11~~~$& $~~~4.51~~~$& $~~~4.17~~~$\\
			\hline
			$~~~R\,({\rm km})~~~$& $~~~19.58~~~$& $~~~19.41~~~$& $~~~19.21~~~$\\
			\hline
		\end{tabular}
		\caption{The maximum mass and corresponding radius for three-form field stiff fluid stars in the presence of a Higgs type potential for $\bar{a}=0.003$, and  different values of $\bar{b}$.}\label{stif-vvar-tab}
	\end{center}
\end{table}

\subsection{Photon stars}

The radiation fluid equation of state $P=\rho/3$ plays a critical role in physics and astrophysics. By using this equation of state we can describe the physical properties of the cores of neutron stars,  assumed to consist of cold, non-interacting and degenerate fermions \cite{Misner:1964zz,R2}.  The interesting possibility of the existence of stars described by the radiation
equation of state, and therefore consisting  of a radiation fluid, was also analyzed from different perspectives \cite{Sorkin:1981wd,Schmidt:1999tr,R5,Chavanis:2007kn}. Einstein’s field equation describing static, spherically symmetric stars made of a radiation fluid, were investigated numerically in \cite{Schmidt:1999tr}.
The  way the thermodynamical parameters (entropy, temperature, baryon number, mass-energy, etc) scale
with the size of a photon fluid star were investigated in \cite{Chavanis:2007kn}, and an unusual behaviour due to the non-extensivity of the system was found. These scaling laws have some similarities with the area scaling law of the black hole entropy.

Another interesting class of stellar-type objects described by the radiation fluid equation of state are represented by the so called ``Radiation Pressure Supported Stars'' (RPSS), which have the intriguing property that they can exist even in classical Newtonian gravity \cite{Mitra:2010ae}. They have been  generalized in order to incorporate the effects of standard general relativity. The corresponding classes of stellar type objects  are called “Relativistic Radiation Pressure Supported Stars” (RRPSS) \cite{Mitra:2010ae}. On the other hand, it was already suggested
in \cite{Glendenning:1997wn} that the formation of RRPSSs may occur during the gravitational collapse of extremely massive baryonic matter clouds,
a process that ends in a very high density state. It turns out that independently of the details of the collapse event, at sufficiently large cosmological redshifts $z\gg 1$, the radiation flux of the collapsed object (star or black hole) always reaches the maximal Eddington luminosity. The properties of the radiation fluid stars in a non-minimally coupled gravity model of the form $Y (R)F^2$, where $F^2$ is the Maxwell invariant and $Y (R)$ is a function of the Ricci scalar $R$, were considered in \cite{Sert:2016kqx}.

In the following we will consider the properties of radiation fluid (photon) stars in three-form field gravity. We will assume again an isotropic pressure mater distribution, with $p_{\perp m}=p_{rm}$, and we consider that the equation of state of the baryonic matter inside the star can be approximated by the photon gas equation of state
\begin{align}
p_{rm}=\frac{\rho_m}{3}.
\end{align}

We will solve numerically the gravitational field equations for central densities that vary
in the range  ${3.1\times10^{14} {\rm g/cm}^3}$ and ${2.9\times 10^{15} {\rm g/cm}^3}$, respectively. We stop the integration when the density reaches the surface value of ${\rho_{mc}= 2\times 10^{14} {\rm g/cm}^3}$ (the nuclear density). In the following, to obtain the interior solutions  we consider
two different choices of the three-form field potential.

\subsubsection{Constant potential: $V(A^2)=\lambda$}

As a first case in the investigation of the photon stars in three-form field gravity we assume that the field potential is a constant,  $V(A^2)=\lambda={\rm constant}$. To numerically integrate the gravitational field equations we consider three different values of $\bar{\lambda}$, $\bar{\lambda}=0,\pm0.02$. For the central density of the photon star we adopt the value $\rho_{mc}=6.4\times 10^{14} {\rm g/cm}^3$.

\begin{figure*}[htbp]
\centering
\includegraphics[width=8.2cm]{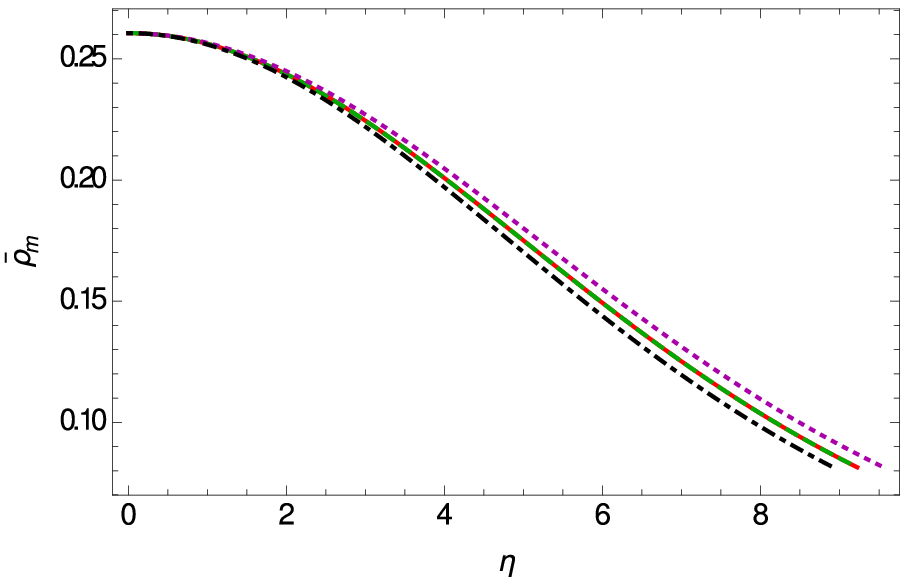}\hspace{.4cm}
\includegraphics[width=8.0cm]{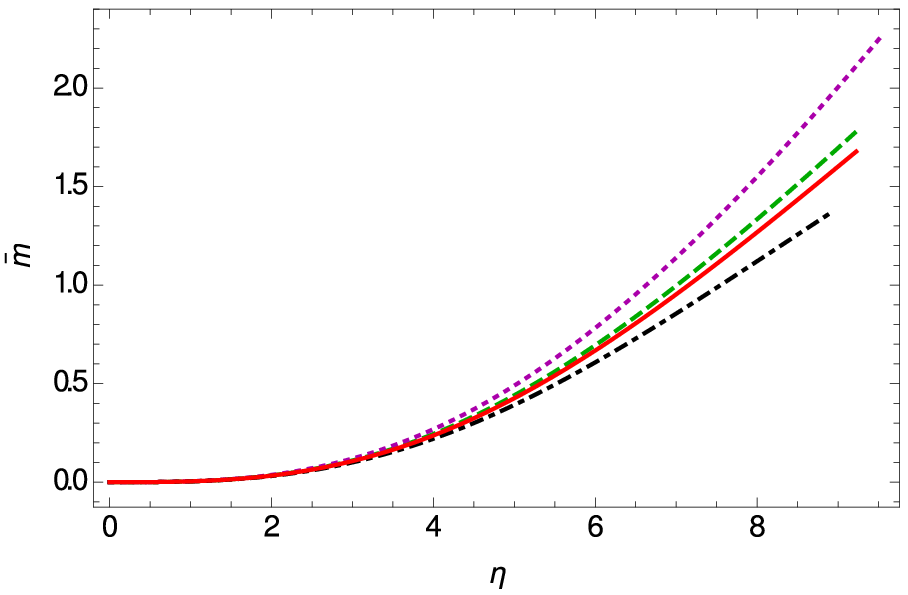}
\caption{Variation of the energy density $\bar{\rho}_{m}$ (left panel) and of the mass $\bar{m}$ (right panel) in terms of $\eta$ for three-form field photon stars in the presence of a constant field potential, for three different values of the constant $\bar{%
\lambda}$: $\bar{\lambda}=0$ (dashed curve),  $\bar{\lambda}=0.02$ (dotted curve), and  $\bar{%
\lambda}=-0.02$ (dot-dashed curve), respectively. The central density of the photon star is taken as $\rho_{mc}=6.4\times10^{14} {\rm g/cm}^3$. The solid curve represents the standard general relativistic density and mass profiles for photon stars.}
\label{fig7}
\end{figure*}

The behaviors of the interior profiles  of the matter energy-density and of the mass are presented in Fig.~\ref{fig7}. The matter energy density is a monotonically decreasing function of the dimensionless radial coordinate, and it reaches the nuclear density at much smaller radial coordinate values, as compared to the stiff fluid case. The matter energy density is also much sensitive to the variations of the potential. The interior mass profile of the star is a monotonically increasing function of $\eta$, and it strongly depends on the numerical values of $\bar{\lambda}$, with the mass of the star increasing for positive values, and decreasing for negative values of $\bar{\lambda}$, respectively.

\begin{figure*}[htbp]
\centering
\includegraphics[width=7.5	cm]{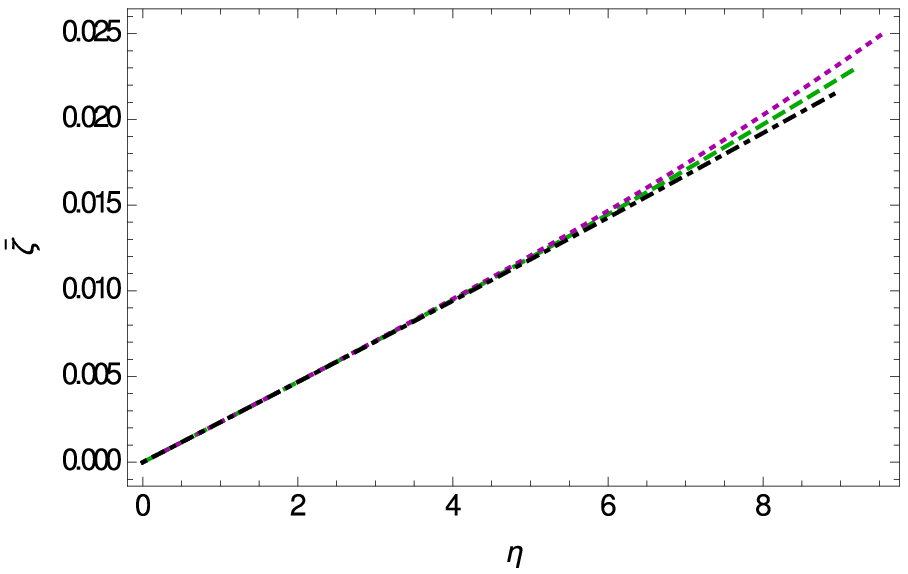}\hspace{0.4cm}
\includegraphics[width=7.5cm]{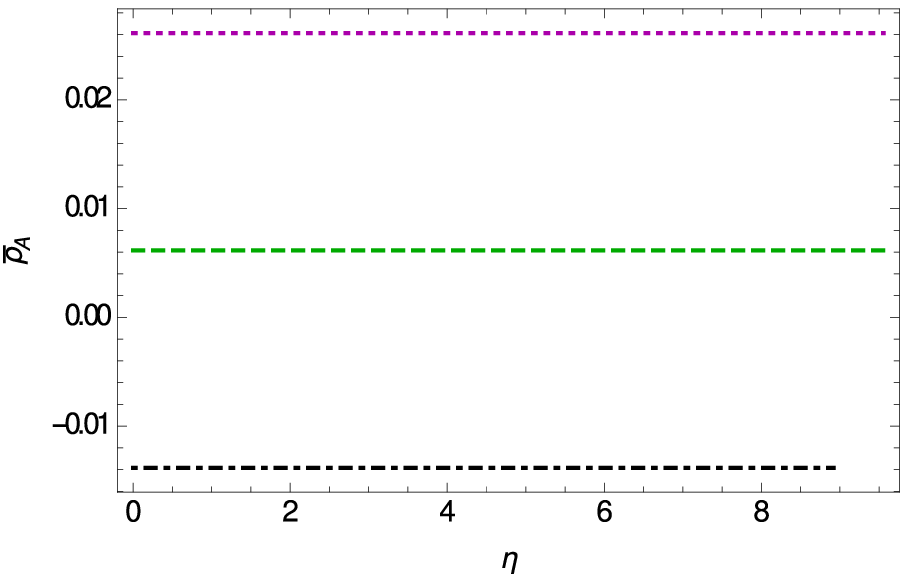}
\caption{Variation of $\bar{\zeta}$(left panel) and $\bar{\rho}_A$ (right panel) as a function  of $\eta$ in three-form field photon stars,
 for three different values of the constant $\bar{\lambda}$: $\bar{\lambda}=0$ (dashed curve),  $\bar{\lambda}=0.02$ (dotted curve), and  $\bar{%
\lambda}=-0.02$ (dot-dashed curve), respectively. The central density of the photon star is taken as $\rho_{mc}=6.4\times10^{14} {\rm g/cm}^3$.}
\label{fig8}
\end{figure*}

The variations of the three-form field $\bar{\zeta}$ and of the effective energy density of the three-form field $\bar{\rho}_A$ in terms of the distance from the center to
the surface of the star are shown, for different values of the constant field potential,  in Fig.~\ref{fig8}.  Similarly to the stiff fluid case, the three-form field is a monotonically increasing function inside the high density object. Near the surface of the star the values of $\bar{\zeta}$ are dependent on the constant values of the field potential. On the other hand, the energy density of the three-form field takes constant values inside the star, and its numerical values are strongly dependent on the model parameters.

\begin{figure}[hh!]
\centering
\includegraphics[width=7.4cm]{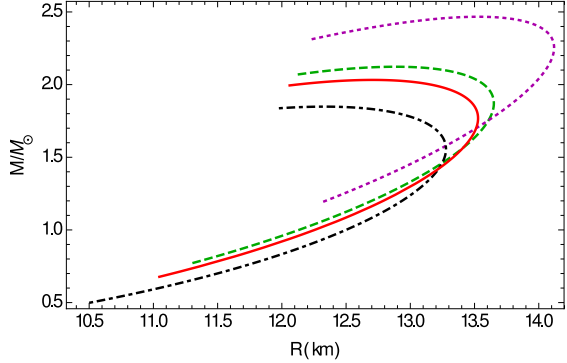}
\caption{Variation of the total mass $M/M_{\odot}$  as a function of the radius of the star $R$ for three-form field photon stars,
for three different values of the constant $\bar{\lambda}$: $\bar{\lambda}=0$ (dashed curve),  $\bar{\lambda}=0.02$ (dotted curve), and  $\bar{%
\lambda}=-0.02$ (dot-dashed curve), respectively. The solid curve represents the mass-radius relation for general relativistic photon stars. }
\label{fig9}
\end{figure}

The mass-radius relation for photon stars is plotted in Fig.~\ref{fig9}. The presence of the three-form field inside the star generates a complicated pattern of behaviors for photon stars. Depending on the sign and numerical value of the constant potential, both higher and lower mass stars do exist in this model. If in standard general relativity the maximum mass of the photon star is close to $2M_{\odot}$, three-form field photon stars can reach maximum masses of around $2.5M_{\odot}$. For negative three-form field potentials smaller masses than in general relativity are also possible. The general shape of the mass-radius function for three-form field photon stars is similar to the standard general relativistic one.

\begin{table}[h!]
\begin{center}
	\begin{tabular}{|c|c|c|c|}
		\hline
 $\bar{\lambda}$ &~~~-$0.02$~~~&$~~~0.0~~~~$&$~~~0.02~~~~$ \\
		\hline
		$~~~\rho_{mc} \times 10^{-14}\,({\rm g/cm}^3)~~~$& $~~~22.8~~~$& $~~~17.0~~~$& $~~~1.23~~~$\\
		\hline
			 \quad$M_{max}/M_{\odot}$\quad& $~~~1.85~~~$& $~~~2.12~~~$& $~~~2.47~~~$\\
		\hline
		 $~~~R\,({\rm km})~~~$& $~~~12.34~~~$& $~~~12.89~~~$& $~~~13.53~~~$\\
		\hline
		\end{tabular}
		\caption{The maximum mass and the corresponding radius for three-form field photon stars with constant potential.}\label{rad-vcons-tab}
	\end{center}
\end{table}

Several numerical values of the maximum mass of photon stars in three-form field gravity are presented in Table~\ref{rad-vcons-tab}. In general relativity for photon stars we have $M_{max}=2.03\,M_{\odot}$, $R=12.71\,{\rm km} $, values corresponding to a central density  $\rho_{mc}=1.87\times 10^{15}\, {\rm g/cm}^3$.

\subsubsection{Higgs-type potential: $V(A^2)=a A^2 +b A^4$}

We consider now the case in which the three-form field photon star also contains a Higgs-type potential, of the form $V(A^2)=a A^2 +b A^4$. For the parameters of the potential we adopt the numerical values $\bar{a}=0.007$ and $\bar{b}=0,\pm 0.11$, respectively, that is, in the following numerical investigations we keep $\bar{a}$ as constant, and we vary $\bar{b}$. Moreover, we fix the central density of the three-form field photon star as $\rho_{mc}=7.12\times 10^{14}\; {\rm g/cm}^3$.

\begin{figure*}[htbp]
\centering
\includegraphics[width=8.1cm]{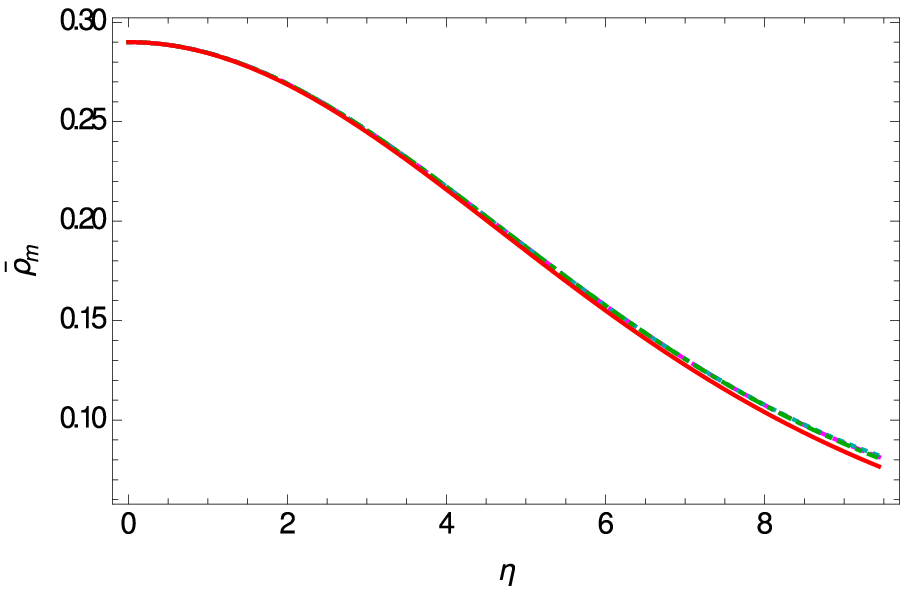}\hspace{.4cm}
\includegraphics[width=8.0cm]{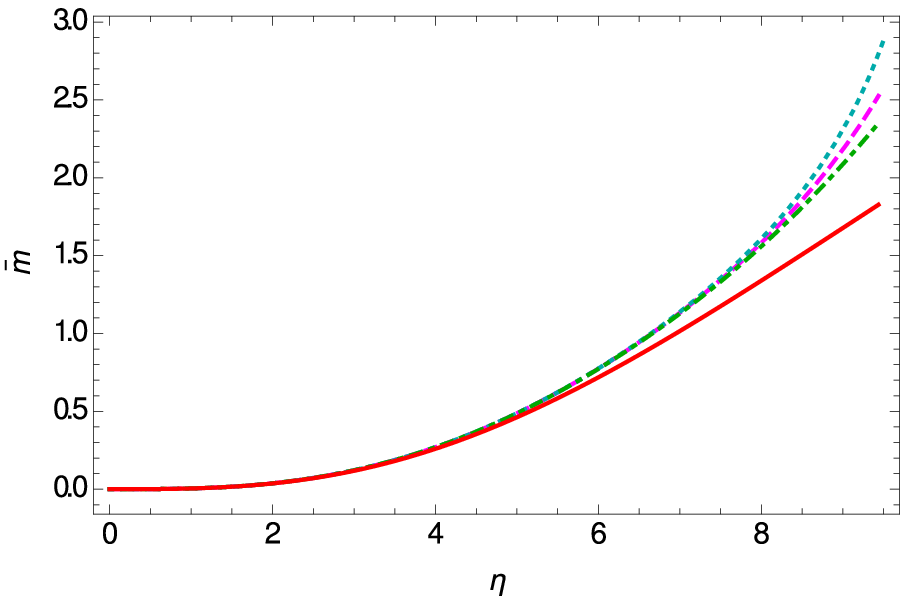}
\caption{Variation of the energy density $\bar{\rho}_m$ (left panel) and of the interior mass profile $\bar{m}$ (right panel) for three-form field photon stars in the presence of a Higgs-type potential as functions of $\eta$, for $\bar{%
a}=0.007$, and for three different values of the constant $\bar{b}$:  $\bar{b}=0$ (dashed curve),  $\bar{b}=-0.11$ (dotted curve), and
$\bar{b}=0.11$ (dot-dashed curve). The central density of the photon star is $\rho_{mc}=7.12\times 10^{14}\; {\rm g/cm}^3$. The solid curve represents the standard general relativistic density and mass profiles for photon stars. }
\label{fig10}
\end{figure*}

\begin{figure*}[htbp]
\centering
\includegraphics[width=8.0cm]{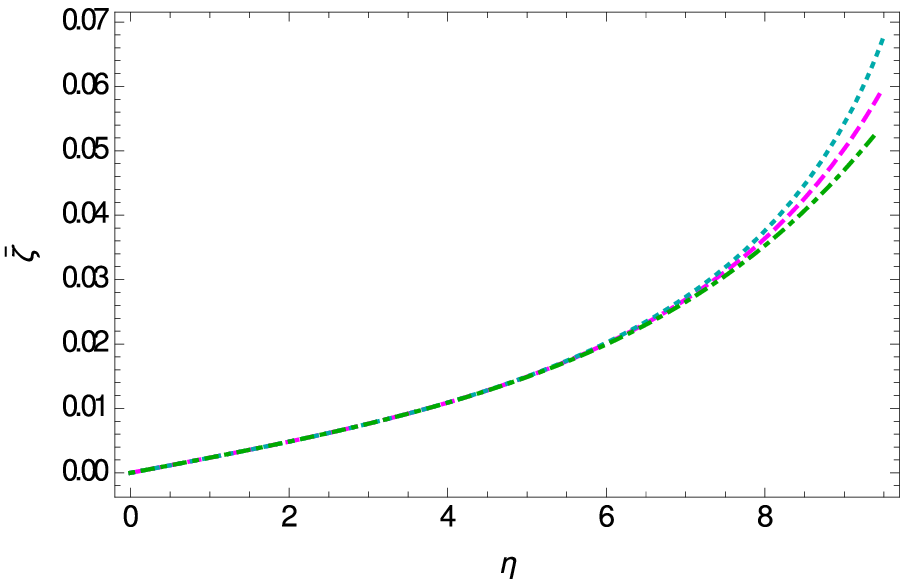}\hspace{.4cm}
\includegraphics[width=8.0cm]{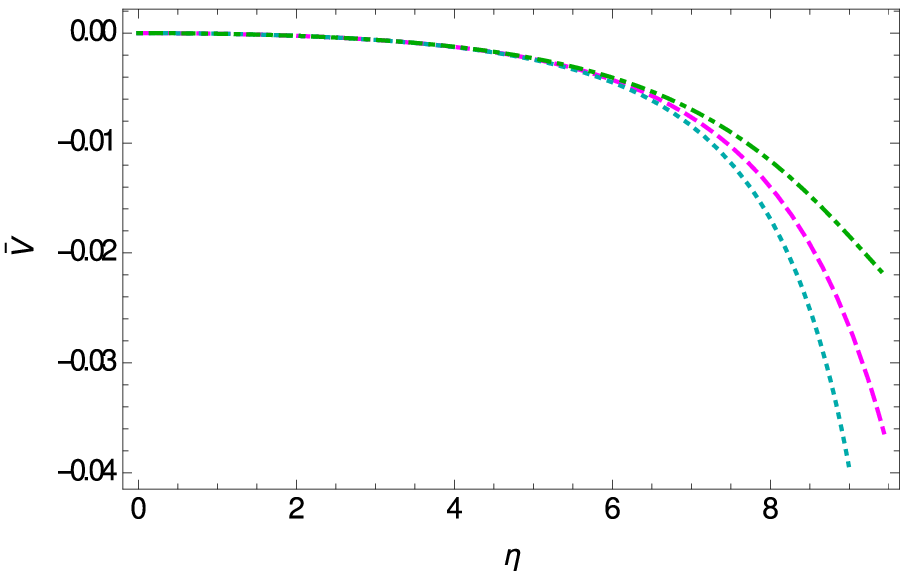}
\caption{Variation of $\bar{\zeta}$ (left panel) and of the Higgs type potential $\bar{V}\left(\zeta\right)$ (right panel) for three-form photon stars  as functions of $\eta$, for $\bar{a}%
=0.007$ and three different values of the constant $\bar{b}$, and for three different values of the constant $\bar{b}$:  $\bar{b}=0$ (dashed curve),  $\bar{b}=-0.11$ (dotted curve), and $\bar{b}=0.11$ (dot-dashed curve), respectively. The central density of the photon star is $\rho_{mc}=7.12\times 10^{14}\; {\rm g/cm}^3$,  while $\bar{\zeta}_0=0$ and $\bar{\zeta}^\prime_0=7\times 10^{-3}$, respectively. }
\label{fig11}
\end{figure*}

The interior energy density and mass profiles of the three-form field photon stars are depicted in Fig.~\ref{fig10}. The energy density is almost insensitive to the variations of the parameters of the Higgs potential. On the other hand, an explicit dependence on the values of $\bar{b}$ exist in the case of the interior mass distribution, which indicates a significant dependence of the mass of the star on the three-form field potential.
We refer the reader to Figs.~ \ref{fig11} for the behavior of the  three-form field component, which is a monotonically increasing function inside the star, reaching its maximum value near the star surface.  Note that the Higgs-type potential is a monotonically decreasing function of $\eta=r/R_0$.

\begin{figure}[htbp]
	\centering
	\includegraphics[width=8.0cm]{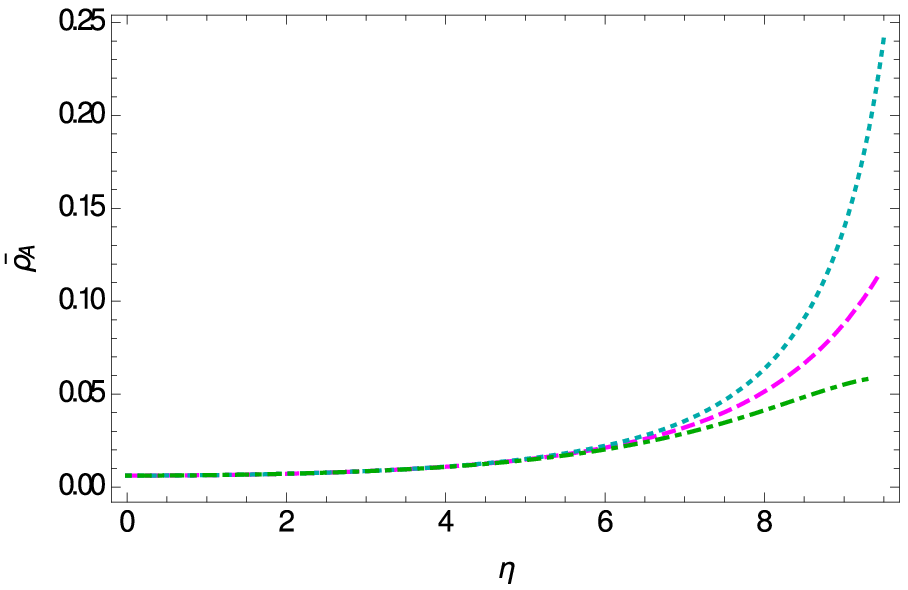}
	\caption{Variation of $\bar{\rho}_A$ for three-form field photon stars  as function of $\eta$, for $\bar{a}%
		=0.007$ and three different values of the constant $\bar{b}$, and for three different values of the constant $\bar{b}$:  $\bar{b}=0$ (dashed curve),  $\bar{b}=-0.11$ (dotted curve), and $\bar{b}=0.11$ (dot-dashed curve), respectively. The central density of the photon star is $\rho_{mc}=7.12\times 10^{14}\; {\rm g/cm}^3$,  while $\bar{\zeta}_0=0$ and $\bar{\zeta}^\prime_0=7\times 10^{-3}$, respectively. }
	\label{densa-rad-vvar}
\end{figure}

The variation of the energy density of the three-form field is presented in Fig.~\ref{densa-rad-vvar}, indicating that $\bar{\rho}_A$ is a monotonically increasing function inside the star.

\begin{figure}[htbp]
\centering
\includegraphics[width=7.5cm]{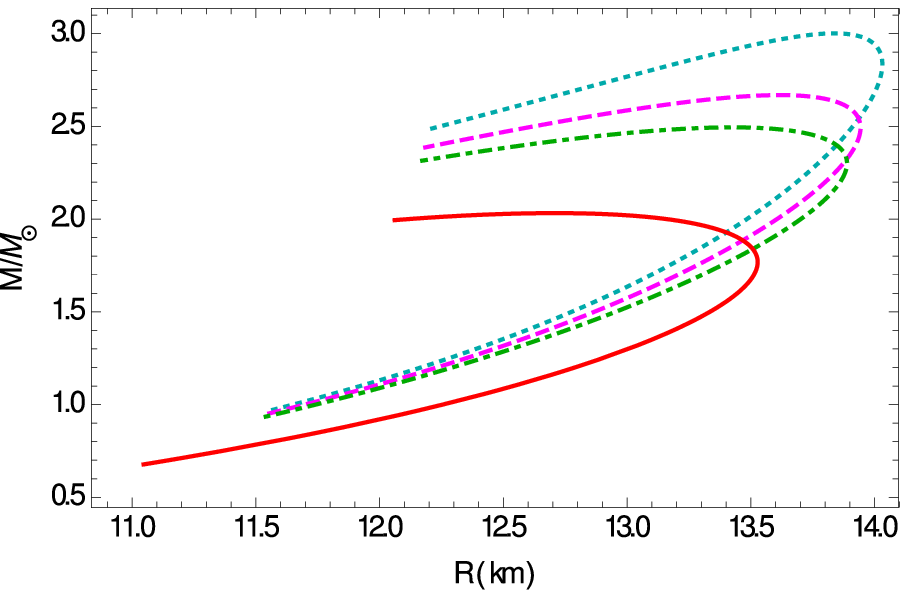}
\caption{Variation of the stellar mass $M/M_{\odot}$ as a function of the radius of the star $R$ for the three-form field photon stars in the presence of a Higgs type potential, for $\bar{a}=0.007$ and three different values of the constant $\bar{b}$: $\bar{b}=0$ (dashed curve),  $\bar{b}=-0.11$ (dotted curve), and $\bar{b}=0.11$ (dot-dashed curve), respectively. The solid curve shows the mass-radius relation for photon stars in standard general relativity.
 }
\label{fig12}
\end{figure}

The mass-radius relation for three-form field photon stars in the presence of a Higgs type potential is presented in Fig.~\ref{fig12}.
As one can see from the $M=M(R)$ relation of three-form field photon stars in the presence of a Higgs field potential, there is a significant increase of the masses of compact objects that essentially depends on the parameters of the potential. If for standard general relativistic photon stars the maximum mass is of the order of $1.9M_{\odot}$, the presence of the Higgs potential leads to an increase of the maximum mass to values of the order of $3M_{\odot}$. There is also a slight increase in the radii of the maximum mass stars, as compared to the standard general relativistic case, which are of the order of 14 km.

\begin{table}[h!]
	\begin{center}
		\begin{tabular}{|c|c|c|c|}
			\hline
			$\bar{b}$ &~~~$-0.11$~~~&$~~~0.0~~~~$&$~~~0.11~~~~$ \\
			\hline
			$~~~\rho_{mc} \times 10^{-14}\,({\rm g/cm}^3)~~~$& $~~~9.53~~~$& $~~~10.3~~~$& $~~~12.5~~~$\\
			\hline
			\quad$M_{max}/M_{\odot}$\quad& $~~~3.00~~~$& $~~~2.67~~~$& $~~~2.49~~~$\\
			\hline
			$~~~R\,({\rm km})~~~$& $~~~13.84~~~$& $~~~13.61~~~$& $~~~13.42~~~$\\
			\hline
		\end{tabular}
		\caption{The maximum masses and the corresponding radii for three-form field photon stars in the presence of a Higgs type potential, for $\bar{a}=0.007$, and  for different values of $\bar{b}$.}\label{rad-vvar-tab}
	\end{center}
\end{table}

A selected set of numerical values of the maximum masses of three-form field photon stars are presented, for different central densities, in Table~\ref{rad-vvar-tab}.

\subsection{Quark stars}

According to the present day view of astrophysics, in neutron stars  nuclear matter, assumed to be in beta equilibrium, is present  from very low densities to several times the nuclear saturation density, given by $\rho _n=0.16\;{\rm fm}^{-3}$ \cite{Shapiro:1983du, Glendenning:1997wn,Graber:2016imq}. However, neutrons are not really elementary particles, and they can be regarded as bound states of their valence quarks and antiquarks. Quarks are spin half particles, and thus are fermions. In the simple phenomenological  MIT bag model, the quarks inside the hadrons are confined by a bag, with the quarks constrained to move freely and independently inside the hadrons by an infinite potential well. In a more general model, called the potential model, quarks are
confined inside the bag by a phenomenological confinement potential, which usually is taken as a harmonic oscillator potential. At the high nuclear densities reached in the interiors of compact neutron stars, a hadron-quark phase transition may occur, leading to the transformation of the neutron star into a quark star \cite{Itoh:1970uw,Bodmer:1971we,Witten:1984rs}.  The equation of state of the quark
matter can be derived from the fundamental Lagrangian of Quantum Chromodynamics (QCD) \cite{Weinberg:1996kr,Mak:2003kw},
\bea
L_{QCD}&=&\frac{1}{4}\sum_{a}F_{\mu \nu }^{a}F^{a\mu \nu }
	\nonumber\\
&&+\sum_{f=1}^{N_{f}}%
\bar{\psi}\left( i\gamma ^{\mu }\partial _{\mu }-g\gamma ^{\mu }A_{\mu }^{a}%
\frac{\lambda ^{a}}{2}-m_{f}\right) \psi ,
\eea
where the subscript $f$ indicates the differen(left panel) and of the Higgs type potential $\bar{V}\left(\zeta\right)$ (right panel)t quark flavors $u$, $d$, $s$, $c$ etc., while the nonlinear gluon field strength is defined according to the Yang-Mills prescription as
$
F_{\mu \nu }^{a}=\partial _{\mu }A_{\nu }^{a}-\partial _{\nu }A_{\mu
}^{a}+gf_{abc}A_{\mu }^{b}A_{\nu }^{c}
$.

An important prediction of QCD is the weakening at short distances of the quark-quark interaction. Under the assumption that  interactions of the quarks and gluons  are weak, for a quark-gluon plasma at
temperature $T$ and having the chemical potential $\mu _{f}$, one can obtain the energy density $\rho $ and the pressure $p$ by using
thermal quantum field theory \cite{Weinberg:1996kr}. By neglecting the quark masses, in the first order perturbation
theory, the equation of state of the quark-gluon plasma is given by \cite{Glendenning:1997wn,Weinberg:1996kr}
\begin{equation}
\varepsilon =\sum_{i=u,d,s,c;e^{-},\mu ^{-}}\varepsilon
_{i}+B,p+B=\sum_{i=u,d,s,c;e^{-}, \mu ^{-}}p_{i},
\end{equation}
where by $B$ (the bag constant) we have denoted the difference between the energy densities of the non-perturbative and perturbative  QCD vacuum. Hence, in the MIT bag model the equation of state for quark matter can be obtained as \cite{Glendenning:1997wn,Weinberg:1996kr,Mak:2003kw}
\begin{equation}
p_{rm}=\frac{1}{3}\left( \rho _{m} -4B\right) .  \label{state}
\end{equation}

Eq.~(\ref{state}) corresponds to the equation of state of a system of massless particles, containing the important corrections originating from the QCD trace anomaly, and from perturbative interactions in the system. All these effects are described by the bag constant $B$. The corrections to the equation of state of the free photon gas are always negative, and they have the important consequence of reducing, at a given temperature,  the energy density of the quark-gluon plasma by a factor of about two \cite{Weinberg:1996kr}. Quark stars have the significant property that their surface density is non-zero, reaching the value $\rho _m=4B$. On the other hand, the quark pressure vanishes on the quark star's surface, thus allowing the definition of the radius of the compact object.

In the following numerical simulations of the structure of isotropic quark stars, with $p_{rm}=p_{\perp m}$, in three-form field gravity we have adopted for the bag constant the numerical value $B=10^{14}\;{\rm g/cm}^3$. We consider central densities in the range $4.2\times 10^{14}\;{\rm g/cm}^3$ and $2.2\times 10^{16}\;{\rm g/cm}^3$, respectively. In all cases, the numerical integration stops at $p_m=0$, which corresponds  $\rho _m=4 B$, according to the equation of state Eq.~\eqref{state}.

\subsubsection{Constant potential case: $V(A^2)=\lambda$}

\begin{figure*}[htbp]
	\centering
	\includegraphics[width=7.8cm]{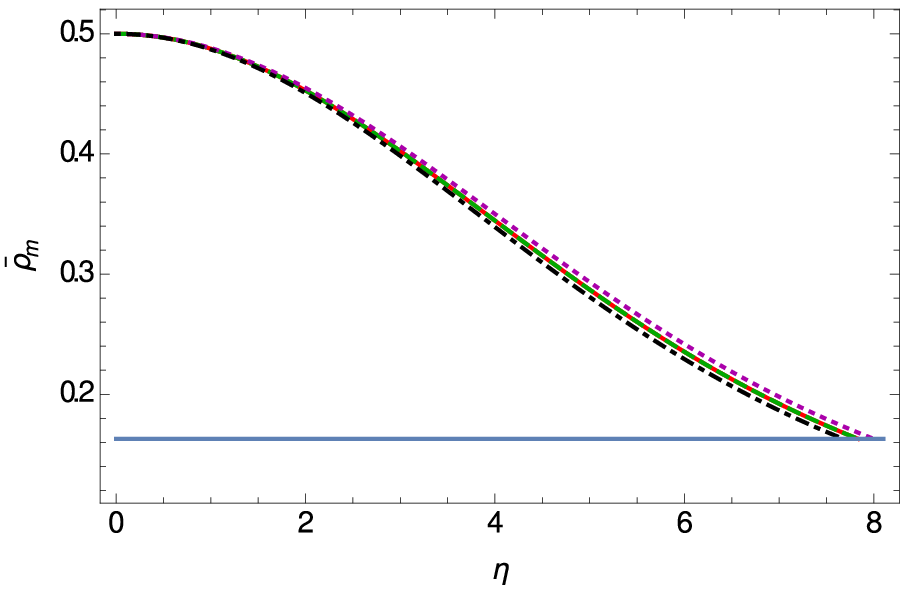}\hspace{.4cm}
	\includegraphics[width=7.8cm]{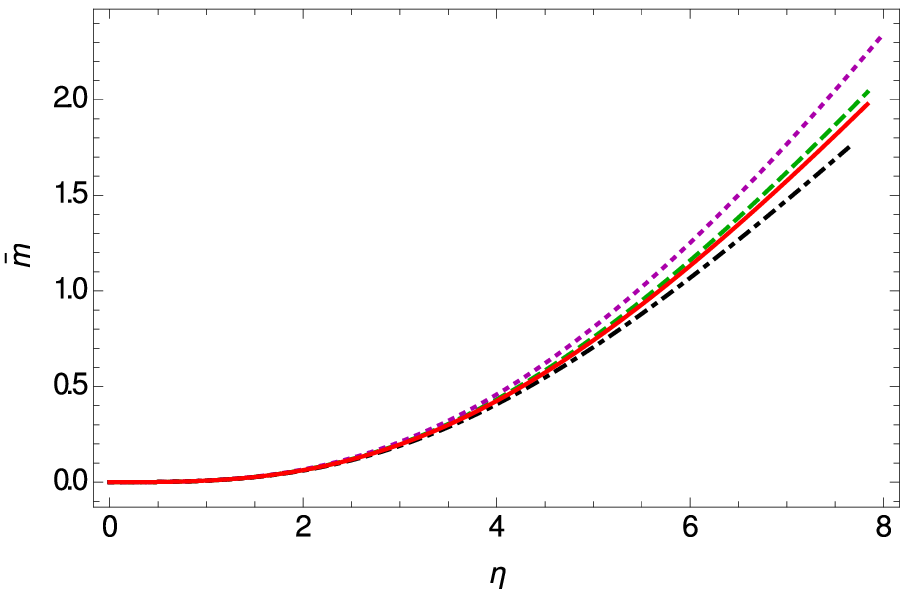}
	\caption{Variation of the energy density $\bar{\rho}_{m}$ (left panel) and of the mass $\bar{m}$ (right panel) in terms of $\eta$ for three-form field quark stars in the presence of a constant field potential, for three different values of the constant $\bar{%
			\lambda}$: $\bar{\lambda}=0$ (dashed curve),  $\bar{\lambda}=0.02$ (dotted curve), and  $\bar{%
			\lambda}=-0.02$ (dot-dashed curve), respectively. The central density of the quark star is taken as $\rho_{mc}=1.23\times10^{15} {\rm g/cm}^3$. The solid curve represents the standard general relativistic density and mass profiles for quark stars. The horizontal line in the left panel shows the surface density $\left.\bar{\rho} _m\right|_{{\rm surface}}=4B$ of the star.}
	\label{dens-mass-quark-vcons}
\end{figure*}

We investigate first quark stars in the presence of a constant three-form field potentia, with $V(A^2)=\lambda$. We consider three different values of $\bar{\lambda}$ as $\bar{\lambda}=0,~\pm 0.02$.
The density and mass profiles for these cases are shown in the Fig.~\ref{dens-mass-quark-vcons}. The density is a monotonically decreasing function of the radial distance, and on the surface of the star it reaches the value $4B$, while the pressure identically vanishes. The density variation is not influenced significantly by the presence of the three-form field, and it is similar to the general relativistic case. The mass profile is monotonically increasing, and indicates that the field potential can have a major influence on the mass of the quark stars.

\begin{figure*}[htbp]
	\centering
	\includegraphics[width=7.8	cm]{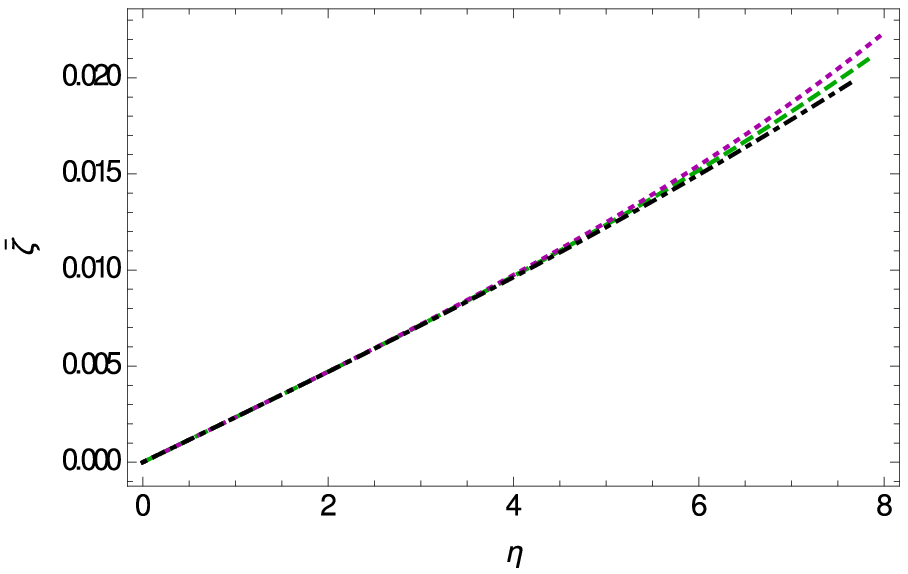}\hspace{.4cm}
	\includegraphics[width=7.8	cm]{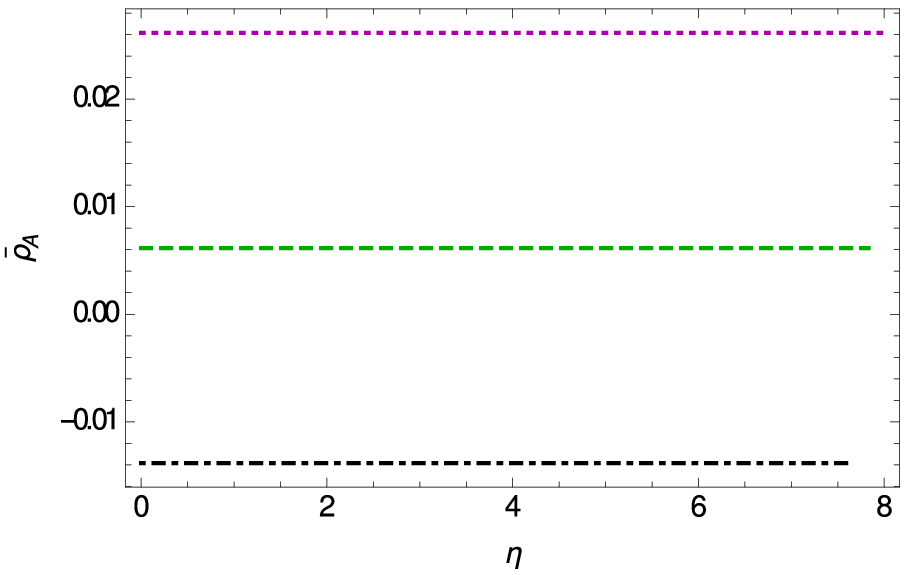}
	\caption{Variation of $\bar{\zeta}$(left panel) and $\bar{\rho}_A$ (right panel)  as a function  of $\eta$ in three-form field quark stars,
		for three different values of the constant $\bar{\lambda}$: $\bar{\lambda}=0$ (dashed curve),  $\bar{\lambda}=0.02$ (dotted curve), and  $\bar{%
			\lambda}=-0.02$ (dot-dashed curve), respectively. The central density of the quark star is taken as $\rho_{mc}=1.23\times10^{15} {\rm g/cm}^3$.}
	\label{quark-xi-vcons}
\end{figure*}

The variations of the three-form field and of $\bar{\rho}_A$ are depicted in the Fig.~\ref{quark-xi-vcons}. Similarly to the previous stellar models, the three-form field is a monotonically increasing function of the radial coordinate $\eta$, reaching its maximum value at the star surface. The variation of the field is slightly dependent on the values of the potential, and this dependence appears mostly near the surface of the star. Similarly to the other constant potential cases, the effective energy density of the three-form field is constant inside the quark star.

\begin{figure}[hh!]
	\centering
	\includegraphics[width=7.4cm]{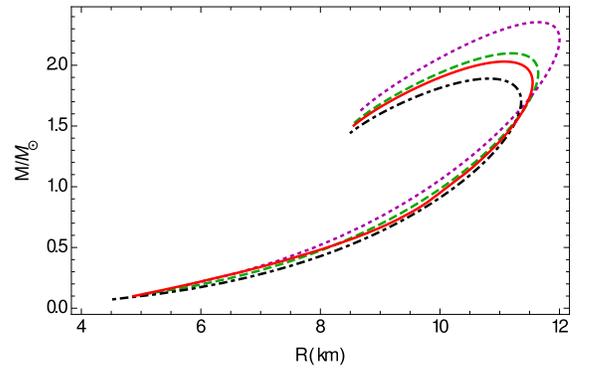}
	\caption{Variation of the total mass $M/M_{\odot}$  as a function of the radius of the star $R$ for three-form field quark stars,
		for three different values of the constant $\bar{\lambda}$: $\bar{\lambda}=0$ (dashed curve),  $\bar{\lambda}=0.02$ (dotted curve), and  $\bar{%
			\lambda}=-0.02$ (dot-dashed curve), respectively. The solid curve represents the mass-radius relation for general relativistic quark stars. }
	\label{quark-mr-vcons}
\end{figure}

 The mass-radius relation for quark stars are shown in Fig.~\ref{quark-mr-vcons}. There is a significant effect of the three-form field, and of its potential, on the maximum masses, and stability regions of quark stars. Increasing the value of the constant field potential leads to a large increase in the maximum mass. However, quark stars with masses smaller than the general relativistic ones are also possible. The general form of the mass-radius relation is also preserved.

The maximum masses of a selected sample of three-form field quark stars are presented, for different central densities,  in Table~\ref{quark-vcons-tab}.

\begin{table}[hh!]
	\begin{center}
		\begin{tabular}{|c|c|c|c|}
			\hline
			$\bar{\lambda}$ &~~~$-0.02$~~~&$~~~0.0~~~~$&$~~~0.02~~~$ \\
			\hline
			$~~~\rho_{mc} \times 10^{-15}\,({\rm g/cm}^3)~~~$& $~~~2.23~~~$& $~~~1.79~~~$& $~~~1.43~~~$\\
			\hline
			\quad$M_{max}/M_{\odot}$\quad& $~~~1.89~~~$& $~~~2.10~~~$& $~~~2.35~~~$\\
			\hline
			$~~~R\,({\rm km})~~~$& $~~~10.82~~~$& $~~~11.20~~~$& $~~~11.64~~~$\\
			\hline
		\end{tabular}
		\caption{The maximum mass and the corresponding radius for three-form field quark stars with constant potential.}\label{quark-vcons-tab}
	\end{center}
\end{table}

In general relativity for quark stars we have ${M_{max}=2.03\,M_{\odot}}$, ${R=11.07\,{\rm km}}$, values corresponding to a central density of ${\rho_{mc}=1.93\times 10^{15}\, {\rm g/cm}^3}$.

\subsubsection{Higgs-type potential: $V(A^2)=a A^2 +b A^4$}

\begin{figure*}[htbp]
	\centering
	\includegraphics[width=8.0cm]{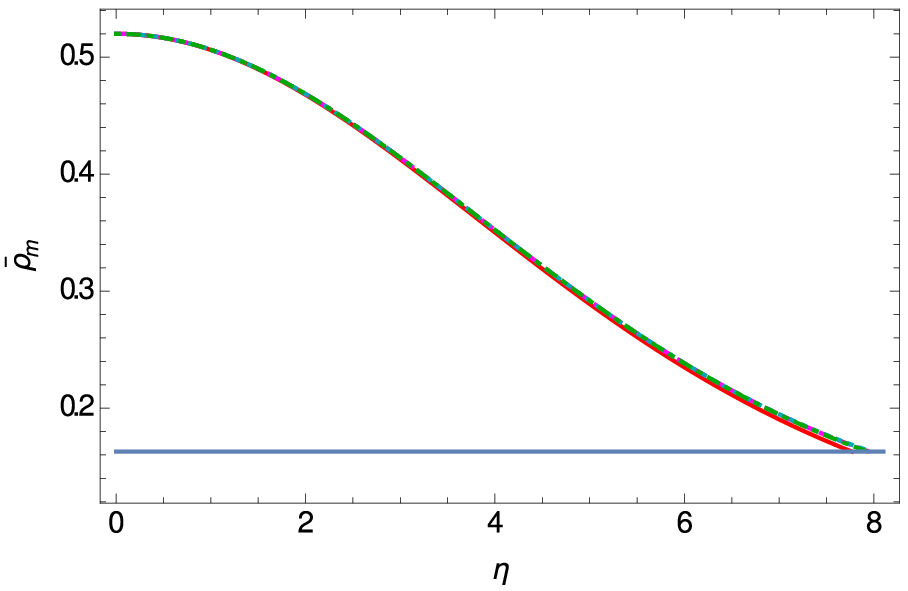}\hspace{.4cm}
	\includegraphics[width=8.0cm]{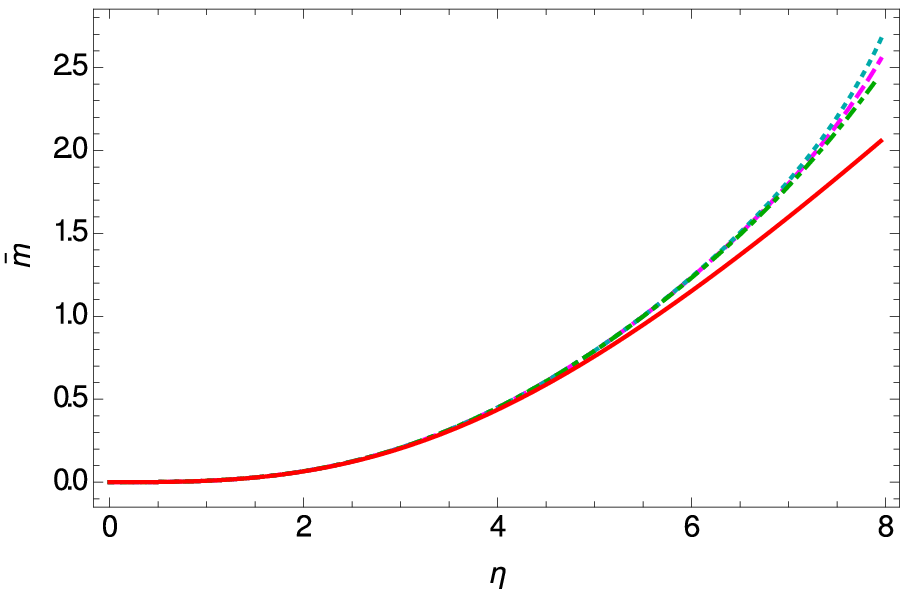}
	\caption{Variation of the energy density $\bar{\rho}_m$ (left panel) and of the interior mass profile $\bar{m}$ (right panel) for three-form field quark stars in the presence of a Higgs type potential as functions of $\eta$, for $\bar{%
			a}=0.01$, and for three different values of the constant $\bar{b}$:  $\bar{b}=0$ (dashed curve),  $\bar{b}=-0.11$ (dotted curve), and
		$\bar{b}=0.11$ (dot-dashed curve). The central density of the quark star is $\rho_{mc}=1.28\times 10^{15}\; {\rm g/cm}^3$. The solid curves correspond to the standard general relativistic case. The horizontal line in the left panel shows the surface density $\left.\bar{\rho} _m\right|_{{\rm surface}}=4B$ of the star.}
	\label{dens-mass-quark-vvar}
\end{figure*}

In the case of three-form field quark stars in the presence of a Higgs-type potential for the model parameters we adopt the values $\bar{a}=0.01$, and we consider three different values of $\bar{b}$, as $\bar{b}=0,~\pm 0.11$.
The density and mass profiles for these forms of the potential  are shown in the Fig. \ref{dens-mass-quark-vvar}. The density monotonically decreases from the center towards the surface of the star, where the pressure vanishes. The radius of the star is not significantly influenced by the parameters of the Higgs potential. However, a significant influence does appear in the case of the mass distribution, with the mass reaching much higher values than in the general relativistic case.

\begin{figure*}[htbp]
	\centering
	\includegraphics[width=8.0cm]{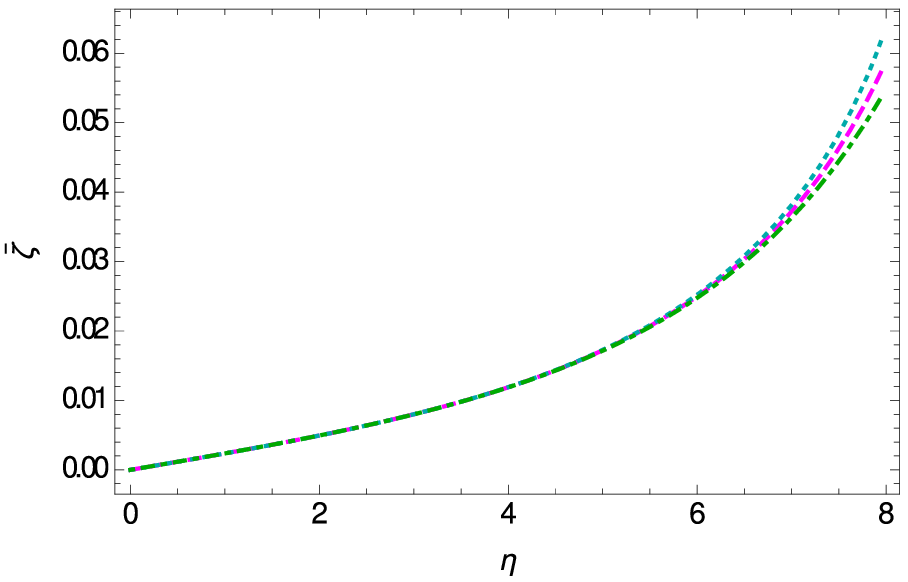}\hspace{.4cm}
	\includegraphics[width=8.1cm]{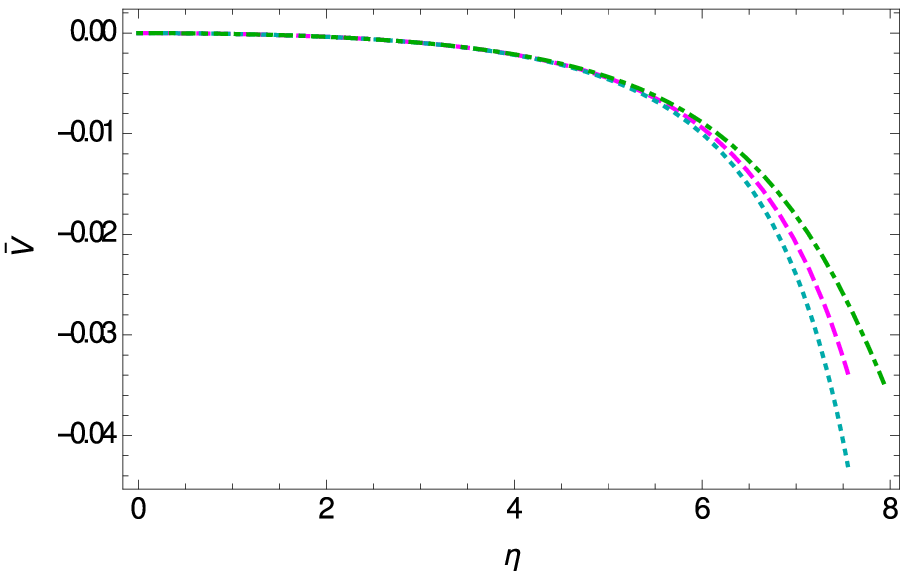}
	\caption{Variation of $\bar{\zeta}$ (left panel) and of the Higgs type potential $\bar{V}\left(\zeta\right)$ (right panel) for three-form quark stars  as functions of $\eta$, for $\bar{a}%
		=0.01$  and for three different values of the constant $\bar{b}$:  $\bar{b}=0$ (dashed curve),  $\bar{b}=-0.11$ (dotted curve), and $\bar{b}=0.11$ (dot-dashed curve), respectively. The central density of the photon star is $\rho_{mc}=1.28\times 10^{15}\; {\rm g/cm}^3$,  while $\bar{\zeta}_0=0$ and $\bar{\zeta}^\prime_0=7\times 10^{-3}$, respectively. }
	\label{xi-poten-quark-vvar}
\end{figure*}

The variation of the three-form field and its Higgs type potential are depicted in the Fig. \ref{xi-poten-quark-vvar}. The three-form field $\bar{\zeta}$ is a monotonically increasing function inside the quark star, reaching its maximum at the surface. Also near the surface the dependence of the field on the parameters of the potential becomes stronger. On the other hand, the Higgs type potential, is a monotonically decreasing curve inside the star, and it takes negative values in all range of distances from the center to the surface. When approaching the surface the variation of the potential shows a significant dependence on the parameters $\bar{a}$ and $\bar{b}$.

\begin{figure}[htbp]
	\centering
	\includegraphics[width=8.0cm]{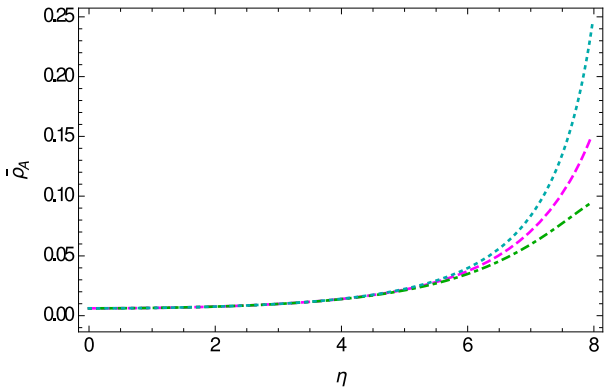}
	\caption{Variation of $\bar{\rho}_A$  for three-form quark stars  as functions of $\eta$, for $\bar{a}%
		=0.01$  and for three different values of the constant $\bar{b}$:  $\bar{b}=0$ (dashed curve),  $\bar{b}=-0.11$ (dotted curve), and $\bar{b}=0.11$ (dot-dashed curve), respectively. The central density of the photon star is $\rho_{mc}=1.28\times 10^{15}\; {\rm g/cm}^3$,  while $\bar{\zeta}_0=0$ and $\bar{\zeta}^\prime_0=7\times 10^{-3}$, respectively. }
	\label{densa-quark-vvar}
\end{figure}

The variation of $\bar{\rho}_A$ for three form quarks stars in the presence of a Higgs type potential is shown in Fig.~\ref{densa-quark-vvar}, indicating an increase of the field energy density inside the quark star.

\begin{figure}[htbp]
	\centering
	\includegraphics[width=8.0cm]{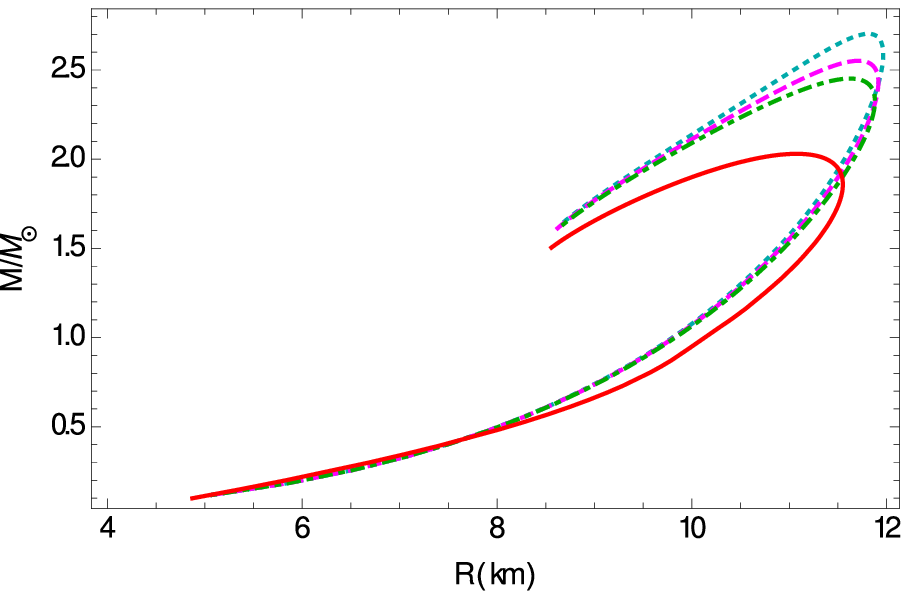}
	\caption{Variation of the stellar mass $M/M_{\odot}$ as a function of the radius of the star $R$ for the three-form field quark stars in the presence of a Higgs type potential, for $\bar{a}=0.01$ and three different values of the constant $\bar{b}$: $\bar{b}=0$ (dashed curve),  $\bar{b}=-0.11$ (dotted curve), and $\bar{b}=0.11$ (dot-dashed curve), respectively. The solid curve shows the mass-radius relation for quark stars in standard general relativity.
	}
	\label{mr-quark-vvar}
\end{figure}

The mass-radius relation for three-form field quark stars in the presence of a Higgs type potential are shown in Fig.~\ref{mr-quark-vvar}. There is also a significant increase of the masses of the quark stars in three-form field gravity, as compared to the general relativistic case. The radii corresponding to the maximum masses also do increase.

Some specific values of the maximum masses of the three-form field quark stars in the presence of the Higgs type potential are presented in Table~\ref{quark-vvar-tab}.

\begin{table}[hh!]
	\begin{center}
		\begin{tabular}{|c|c|c|c|}
			\hline
			$\bar{b}$ &~~~$-0.11$~~~&$~~~0.0~~~~$&$~~~0.11~~~~$ \\
			\hline
			$~~~\rho_{mc} \times 10^{-15}\,({\rm g/cm}^3)~~~$& $~~~1.22~~~$& $~~~1.29~~~$& $~~~1.36~~~$\\
			\hline
			\quad$M_{max}/M_{\odot}$\quad& $~~~2.70~~~$& $~~~2.55~~~$& $~~~2.45~~~$\\
			\hline
			$~~~R\,({\rm km})~~~$& $~~~11.82~~~$& $~~~11.73~~~$& $~~~11.65~~~$\\
			\hline
		\end{tabular}
		\caption{The maximum masses and the corresponding radii for three-form field quark stars in the presence of a Higgs type potential for $\bar{a}=0.01$ and  different values of $\bar{b}$.}\label{quark-vvar-tab}
	\end{center}
\end{table}

\subsection{Bose-Einstein Condensate stars}

If the temperature $T$ of the boson gas drops below the critical value $T_{cr}$, given by ${T_{cr}=2\pi\hbar^2\rho_{cr}^{2/3}/ \zeta^{2/3}(3/2)m^{5/3}k_B}$, where $m$ is the particle mass in the condensate, $\rho_{cr}$ is the
critical transition density, $k_B$ is Boltzmann's constant, and $\zeta $ is the
Riemann zeta function, respectively, then the gas undergoes a phase transition becoming a Bose-Einstein Condensate, with all particles in the system quantum mechanically correlated \cite{BEC1}. It is assumed that due to the superfluid properties of the neutron matter a significant amount of mass of astrophysical compact objects may be in the form of a Bose-Einstein Condensate \cite{Glendenning:1997wn, Chavanis:2011cz}. In the non-relativistic and the Newtonian theoretical limits a Bose-Einstein Condensate confined gravitationally can be described as a gas, with the pressure and density related by a
barotropic equation of state of the form $p=p(\rho)$. The equation of state of the condensate is characterized by two important physical
parameters, the mass of the condensate particle $m$, and the scattering length $a$ \cite{Boehmer:2007um}. For a condensate with
quartic non-linearity, the equation of state is polytropic with index $n = 1$, and it is given by $p (\rho)=K\rho ^2$ \cite{Boehmer:2007um,Harko:2019nyw,Craciun:2020twu}. In the case of a neutron condensate fluid, in which the nuclear matter underwent a phase transition after the neutrons formed Copper pairs,  the constant $K$ is given by \cite{Chavanis:2011cz}
\begin{equation}
K=\frac{2\pi \hbar ^{2}a}{m_c^{3}} =0.1856\times 10^5 \left(\frac{a}{1\;%
\mathrm{fm}}\right)\left(\frac{m_c}{2m_n}\right)^{-3},
\end{equation}
where $m_n=1.6749\times 10^{-24}$ g denotes the mass of the neutron. If the particles at the core of compact high
density stellar objects form Cooper pairs having masses of the order of two neutron masses, and by assuming a
scattering length of the order of 10-20 fm, Bose-Einstein Condensate stars can achieve maximum central densities of the order
of $0.1-0.3 \times 10^{16}$ g/cm$^3$, minimum radii in the range of 10-20 km, and maximum masses of the order of 2$M_{\odot}$ \cite{Chavanis:2011cz}.

In the following we will investigate the properties of Bose Einstein Condensate/superfluid stars in three-form field gravity theory. We assume that matter inside the star can be described by the equation of state
\begin{align}
p_{rm}=K \rho_m^\gamma,
\end{align}
where $\gamma $ is a constant. We consider again an isotropic pressure distribution with $p_{\perp m}=p_{rm}$. We consider that the superfluid interior of the star can be described by a Bose-Einstein Condensate, and hence we adopt for $\gamma$  the value $\gamma=2$.
As for the constant $K$ in the following we fix it to the rescaled value $\bar{K}=\epsilon_0 K=0.4$. We will numerically investigate three-form field Bose-Einstein Condensate stars for two different choices of the field potential. In our analysis we assume that the central density varies between
the values $2.44\times 10^{14}\; {\rm g/cm}^3$ and $6.43\times 10^{15}\; {\rm g/cm}^3$, respectively.

\subsubsection{Constant potential: $V(A^2)=\lambda$}

\begin{figure*}[htbp]
\centering
\includegraphics[width=8.1cm]{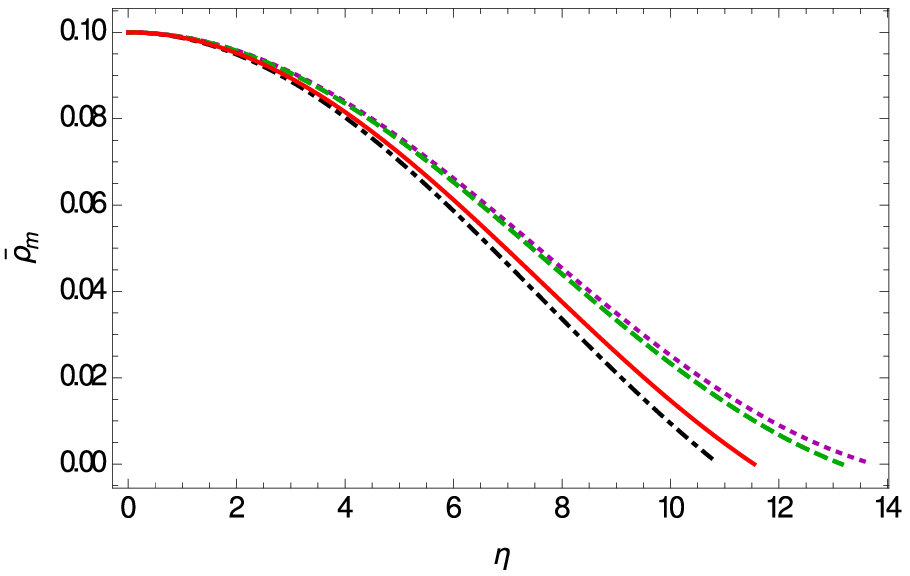}\hspace{.4cm}
\includegraphics[width=8.0cm]{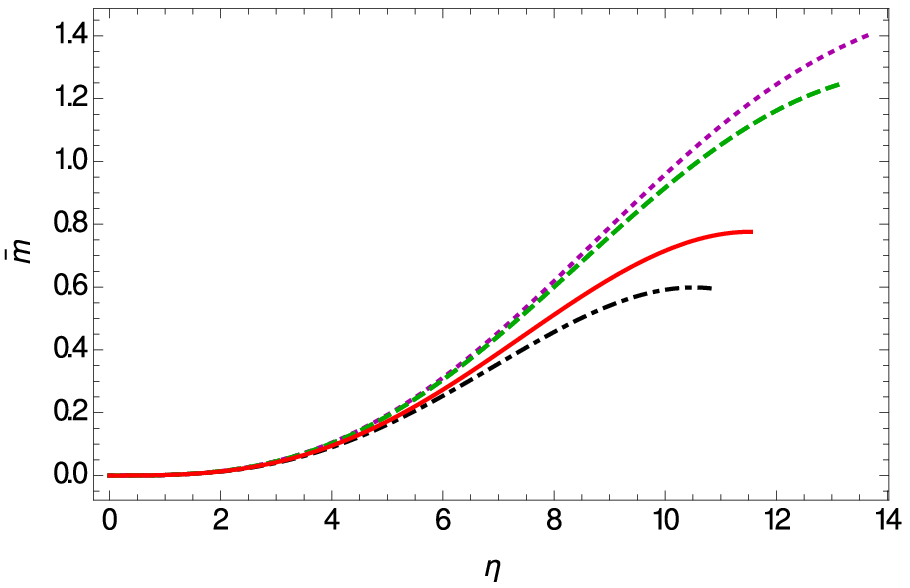}
\caption{Variations of the density profiles $\bar{\rho}_m $ of the superfluid matter (left panel) and of the normalized mass profile $\bar{m}$ (right panel) as a function of the radial distance $\eta$ for the three-form field Bose-Einstein Condensate stars, for three different values of the
constant $\bar{\lambda}$:  $\bar{%
\lambda}=0$ (dashed curve),  $\bar{\lambda}=0.0013$ (dotted curve), and
 $\bar{\lambda}=-0.01$ (dot-dashed curve). The solid curves represent the corresponding standard general relativistic quantities.
 }
\label{fig13}
\end{figure*}

As a first example of a Bose-Einstein Condensate star in three-form field gravity we consider the case of the constant field potential, $V(A^2)=\lambda={\rm constant}$. We will adopt three different choices for $\bar{\lambda}$ as $\bar{\lambda}=0,
\,0.0013$ and $-0.01$, respectively. The variation of the matter density and of the mass profile inside the three-form field Bose-Einstein Condensate star are represented in Fig.~\ref{fig13}.

 As one can see from the plots, the density profile is a monotonically decreasing function of the radial coordinate, while the mass profile is a monotonically increasing function of $\eta=r/R_0$. The density, as well as the pressure vanish on the vacuum boundary of the star, and hence the radius of the star is uniquely defined. Both the density and the mass profiles show a significant dependence on the numerical values of the potential, and a significant increase in the numerical values of the mass of the condensate star can be obtained by varying the values of $\bar{\lambda}$.

\begin{figure*}[htbp]
\centering
\includegraphics[width=7.8cm]{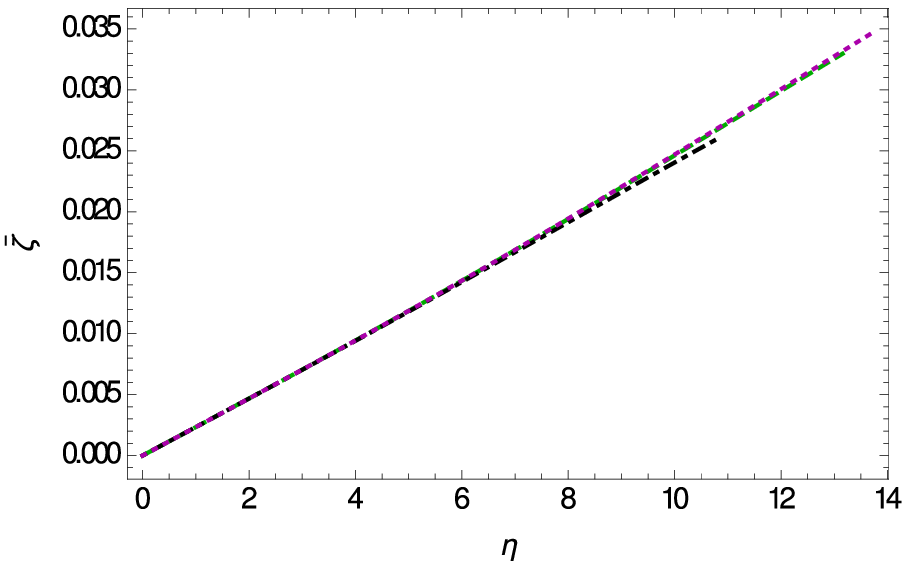}\hspace{0.4cm}
\includegraphics[width=7.8cm]{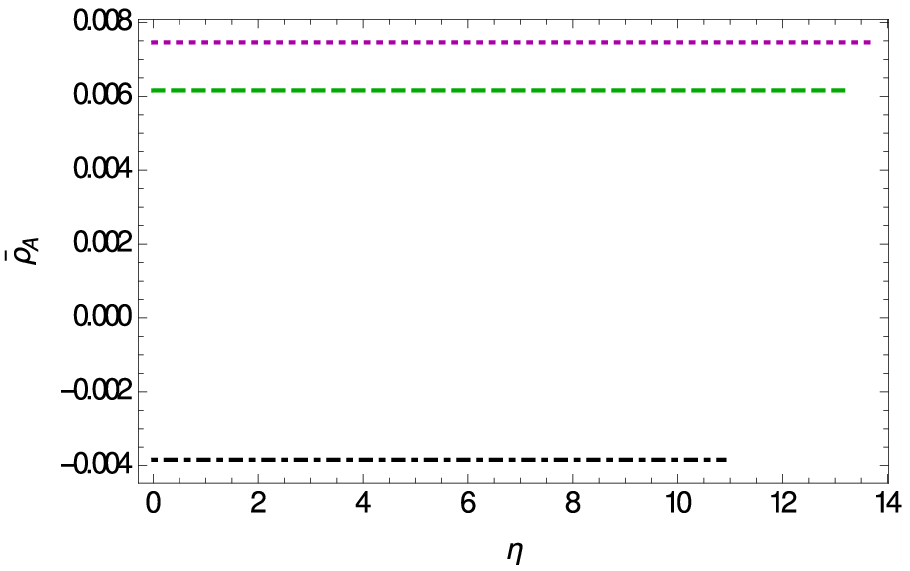}
\caption{Variation of $\bar{\xi}$ (left panel) and $\bar{\rho}_A$ (right panel)
as a function of distance from the center of the star $\eta$ for the three-form field Bose-Einstein Condensate stars, for three different values
of the constant field potential $\bar{\lambda}$: $\bar{%
\lambda}=0$ (dashed curve),  $\bar{\lambda}=0.0013$ (dotted curve), and
 $\bar{\lambda}=-0.01$ (dot-dashed curve).}
\label{fig14}
\end{figure*}

The variations of the non-zero component of the three-form field and of the energy density $\bar{\rho}_A$ in the interior of the star are depicted in Fig.~\ref{fig14}.
The three-form field is a monotonically increasing function of the radial distance from the center of the star, and it reaches its maximum on the star surface. For the adopted set of potential values the variation of $\bar{\zeta}$ is almost independent on $\bar{\lambda}$. The energy density of the field remains a constant inside the star.

\begin{figure}[htbp]
\centering
\includegraphics[width=7.8cm]{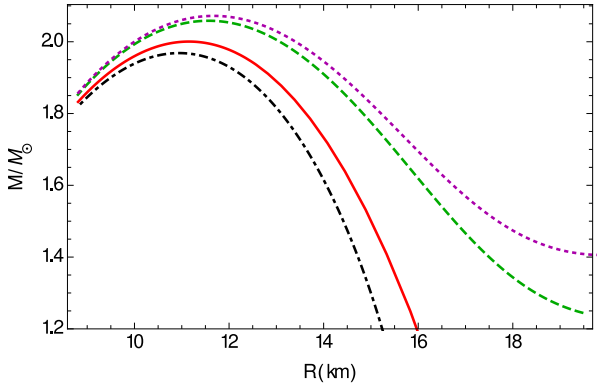}
\caption{Variations of the mass $M/M_\odot$ of the three-form field Bose-Einstein Condensate stars as functions of the radius of the star $R$, for three different values
of the constant field potential $\bar{\lambda}$: $\bar{%
\lambda}=0$ (dashed curve),  $\bar{\lambda}=0.0013$ (dotted curve), and
 $\bar{\lambda}=-0.01$ (dot-dashed curve). The solid curve represents the mass-radius relation for Bose-Einstein Condensate stars in standard general relativity.}
\label{fig15}
\end{figure}

The mass-radius relation of the three-form field Bose-Einstein Condensate  stars is depicted in
Fig.~\ref{fig15}.
The three-form field Bose-Einstein Condensate stars with a constant field potential may have higher masses than the similar general relativistic stars. The mass increase/decrease depends on the sign of the constant potential. The increase of the mass is also associated to a slight increase of the radius of the star. A selected number of numerical values of the maximum masses of the three-form field Bose-Einstein Condensate stars in the presence of a constant potential are presented in Table~\ref{BE-vcons-tab}.

\begin{table}
\begin{center}
	\begin{tabular}{|c|c|c|c|}
		\hline
		$\bar{\lambda}$ &~~~-$0.01$~~~&$~~~0.0~~~~$&$~~~0.0013~~~$ \\
		\hline
		$~~~\rho_{mc} \times 10^{-15}\,({\rm g/cm}^3)~~~$& $~~~2.77~~~$& $~~~2.29~~~$& $~~~2.21~~~$\\
		\hline
		\quad$M_{max}/M_{\odot}$\quad& $~~~1.97~~~$& $~~~2.06~~~$& $~~~2.07~~~$\\
		\hline
		$~~~R\,({\rm km})~~~$& $~~~10.93~~~$& $~~~11.56~~~$& $~~~11.68~~~$\\
		\hline
	\end{tabular}
	\caption{The maximum masses and the corresponding radii for three-form field Bose-Einstein stars with constant potential.}\label{BE-vcons-tab}
\end{center}
\end{table}

In general relativity for Bose-Einstein Condensate stars we have $M_{max}=2.00\,M_{\odot}$, $R=11.15\,{\rm km} $, values corresponding to a central density ${\rho_{mc}=2.58\times 10^{15}\, {\rm g/cm}^3}$.

\subsubsection{Higgs-type potential: $V(A^2)=a A^2 +b A^4$}

We consider now three-form field Bose-Einstein Condensate stars in the presence of a Higgs-type potential $V(A^2)=a A^2 +b A^4$. For our numerical investigations we fix the value of $\bar{a}$ as $\bar{a}=0.001$, and we consider three different choices of $\bar{b}$
as $\bar{b}=0,\, 0.6$ and $-0.1$. In the cases $\bar{b}=0,\, 0.6$ the
density profile has a local minimum,  where its value is still greater than zero.
Since the density profile should be a decreasing function of the distance
from the center of the compact star, we stop integration around the minimum density $%
2.40\times 10^{13}\, {\rm g/cm}^3$, which is smaller than the nuclear density. However, in the case $\bar{b}=-0.1$,  the density vanishes at a finite $r/R_0$, and hence the radius of the star can be determined exactly.

\begin{figure*}[htbp]
\centering
\centering\includegraphics[width=7.8cm]{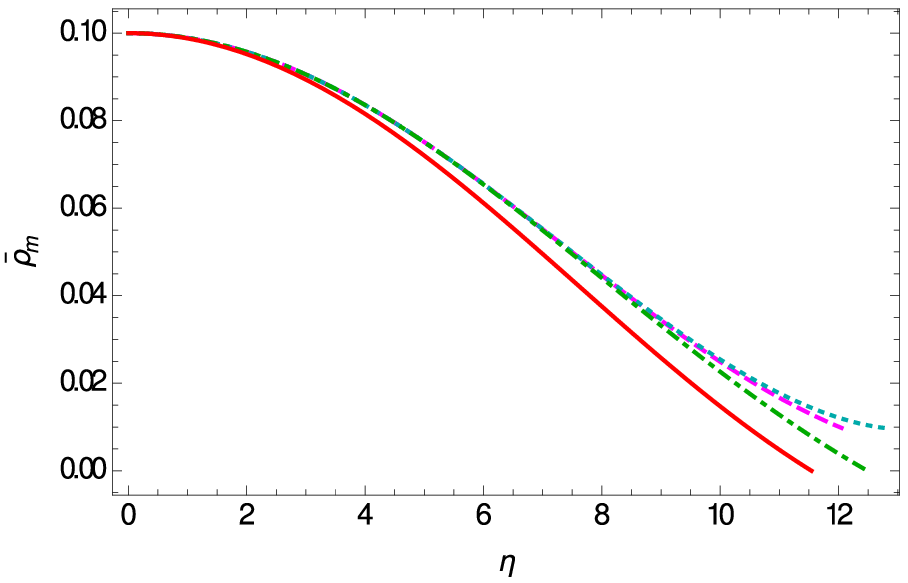}\hspace{.4cm}
\includegraphics[width=8.0cm]{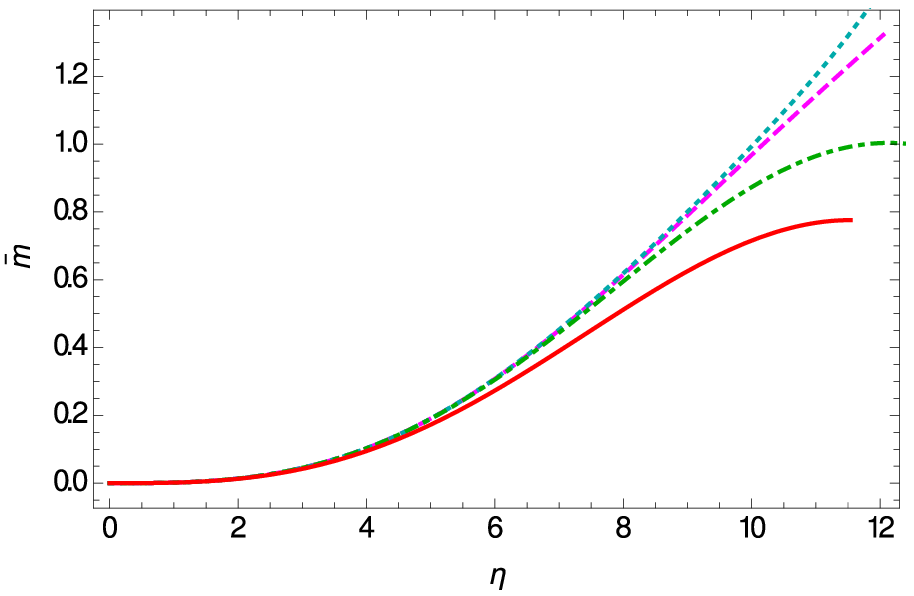}
\caption{Variations of the matter density profile $\bar{\rho}_m$ (left panel) and of the mass profile $\bar{m}$ as functions of $\eta$ for the three-form field Bose-Einstein Condensate stars in the presence of a Higgs type potential, for $\bar{a}=0.001$ and
three different values of the constant $\bar{b}$: $\bar{b}=0$ (dashed curve), $\bar{b}=-0.1$ (dotted curve),
and $\bar{b}=0.6$  (dot-dashed curve), respectively.  The solid represents the same quantities in standard general relativity. }
\label{fig16}
\end{figure*}

The variations of the densities and masses of the three-form field Bose-Einstein Condensate stars in the presence of a Higgs potential are plotted in  Fig.~\ref{fig16}. The density of the condensate stars is a monotonically decreasing function of $rR_0$, but its behavior near the star surface strongly depends on the model parameters. The radius of the star can be defined either exactly, as the radial distance at which the density vanishes, or as the point corresponding to the local minimum of the density. Once the cut-off point of the density is obtained, one can also uniquely find the maximum mass of the star.

\begin{figure*}[htbp]
\centering
\includegraphics[width=8.0cm]{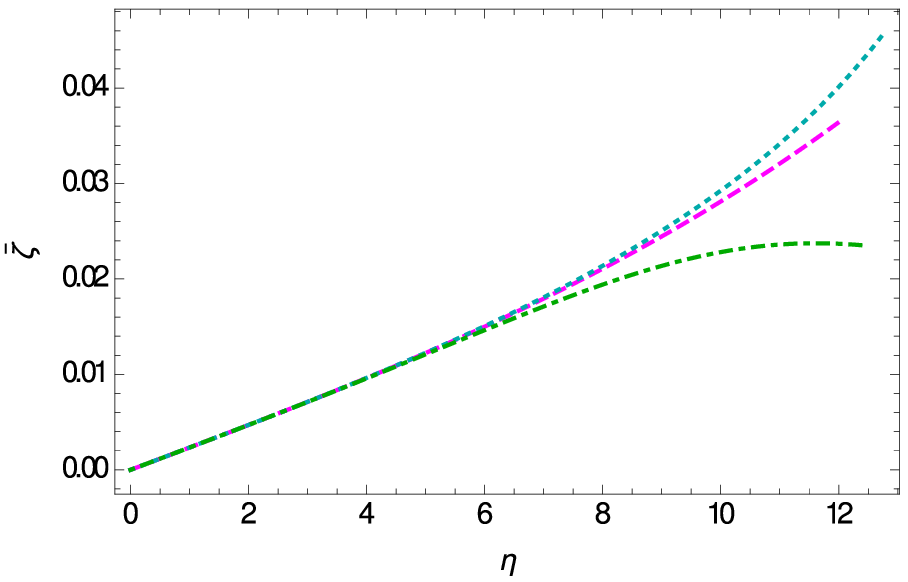}\hspace{.4cm}
\includegraphics[width=8.0cm]{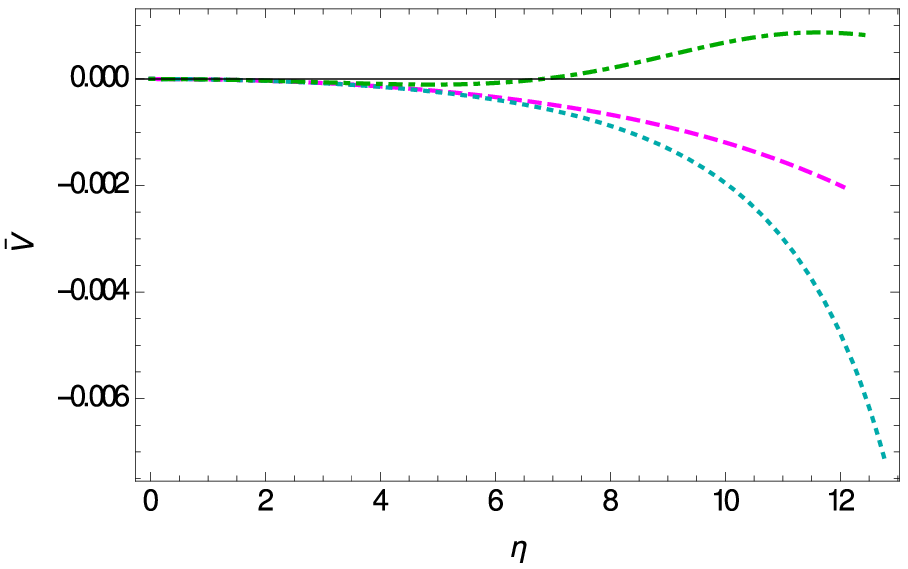}
\caption{Variation of the non-zero component of the three-form field $\bar{\zeta}$ (left panel) and of the Higgs type potential (right panel) as function of the radial coordinate $\eta$ for three-form field Bose-Einstein Condensate stars in the presence of a Higgs type potential for
$\bar{a}=0.001$, and three different values of the constant $\bar{b}$: $\bar{b}=0$ (dashed curve), $\bar{b}=-0.1$ (dotted curve),
and $\bar{b}=0.6$  (dot-dashed curve), respectively. }
\label{fig17}
\end{figure*}

The variations of the three-form field component $\bar{\zeta}$ and of the Higgs type potential $\bar{V}\left(\bar{\zeta}\right)$ are represented in Fig.~\ref{fig17}. While for smaller values of $\bar{b}$, $\bar{\zeta}$ is a monotonically decreasing function, reaching its maximum on the stellar surface, for higher values of $\bar{b}$, $\bar{\zeta}$ reaches its maximum below the surface, and tends to decrease towards the vacuum boundary. The dependence of the behavior of the potential on $\bar{b}$ is even stronger. For large values of $\bar{b}$ the potential becomes a positive function with the maximum beyond the star's surface, while for smaller values it is a monotonically decreasing function of the radial coordinate, taking negative values inside the star.

\begin{figure}[htbp]
	\centering
	\includegraphics[width=8.0cm]{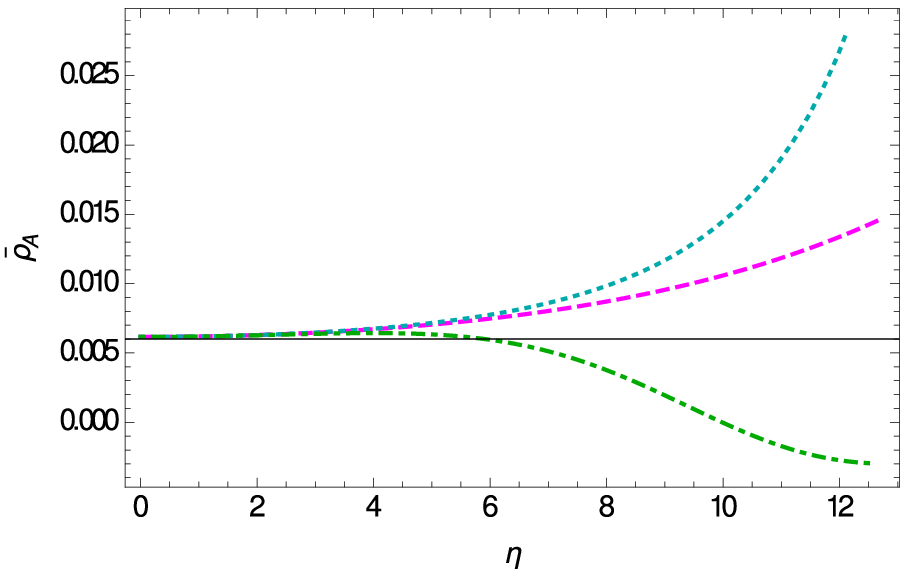}
	\caption{Variation of the energy  density of the three-form field $\bar{\rho}_A$  as function of the radial coordinate $\eta$ for three-form field Bose-Einstein Condensate stars in the presence of a Higgs type potential for
		$\bar{a}=0.001$, and three different values of the constant $\bar{b}$: $\bar{b}=0$ (dashed curve), $\bar{b}=-0.1$ (dotted curve),
		and $\bar{b}=0.6$  (dot-dashed curve), respectively. }
	\label{densa-BE-vvar}
\end{figure}

The variation of $\bar{\rho}_A$ inside the Bose-Einstein Condensate star is represented in Fig.~\ref{densa-BE-vvar}. The energy density of the field has a complex behavior, which is strongly dependent on the model parameters, and it can take both positive and negative values inside the star.

\begin{figure}[htbp]
\centering
\includegraphics[width=8.0cm]{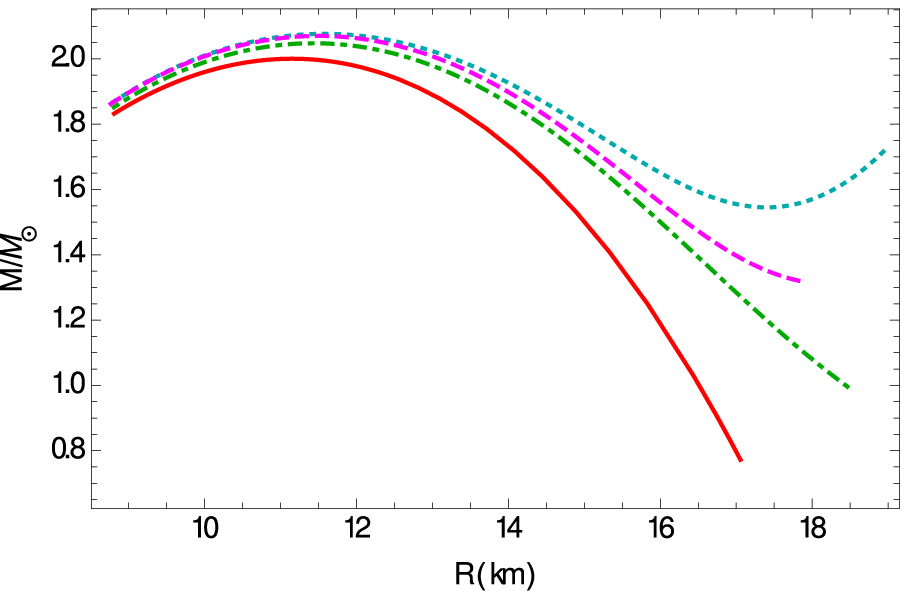}
\caption{Variation of $M/M_{\odot}$ as a function of the radius $R$ for three-form field Bose-Einstein Condensate stars in the presence of a Higgs type potential, for $\bar{a}%
=0.001$ and three different values of the constant $\bar{b}$: $\bar{b}=0$ (dashed curve), $\bar{b}=-0.1$ (dotted curve),
and $\bar{b}=0.6$  (dot-dashed curve), respectively. The solid curve represents the mass-radius relation for general
relativistic Bose-Einstein Condensate stars. }
\label{fig18}
\end{figure}

The mass-radius relations for three-form field Bose-Einstein Condensate stars in the presence of a Higgs type potential are depicted in Fig.~\ref{fig18}.
Three-form field Bose-Einstein Condensate stars in the presence of a Higgs potential can reach higher maximum masses as compared to the general relativistic case, which can exceed $2M_{\odot}$. However, no significant variation in the radii of the stars does appear due to the presence of the three-form field. Specific values of the maximum masses of three-form field Bose-Einstein Condensate stars in the presence of a Higgs type potential are presented in Table~\ref{BE-vvar-tab}.

\begin{table}[h!]
	\begin{center}
		\begin{tabular}{|c|c|c|c|}
			\hline
			$\bar{b}$ &~~~$-0.1$~~~&$~~~0.0~~~~$&$~~~0.6~~~~$ \\
			\hline
			$~~~\rho_{mc} \times 10^{-15}\,({\rm g/cm}^3)~~~$& $~~~2.20~~~$& $~~~2.21~~~$& $~~~2.37~~~$\\
			\hline
			\quad$M_{max}/M_{\odot}$\quad& $~~~2.08~~~$& $~~~2.07~~~$& $~~~2.04~~~$\\
			\hline
			$~~~R\,({\rm km})~~~$& $~~~11.56~~~$& $~~~11.50~~~$& $~~~11.44~~~$\\
			\hline
		\end{tabular}
		\caption{The maximum masses and the corresponding radii for three-form field Bose-Einstein Condensate stars in the presence of a Higgs type potential for $\bar{a}=0.001$, and  different values of $\bar{b}$.}\label{BE-vvar-tab}
	\end{center}
\end{table}

\section{Discussions and final remarks}\label{conclusions}

In the present paper, we have investigated the possible  existence of stellar-type massive compact astrophysical objects, described, together with their baryonic content, by the three-form field gravitational theory, in which the standard Hilbert-Einstein action of general relativity is non-trivially extended by the addition of the Lagrangian of a three-form field to the total action. In order to investigate the gravitational properties of this theoretical  model we have considered the simplest case, corresponding to a stellar interior described by a static and spherically symmetric geometry. In this case the system of the gravitational field equations in three-form field gravity also depends on the scalar radial function $\zeta$, the radial component of the dual vector $B^{\delta}$ of the three-form $A_{\alpha \beta \gamma}$, and on the arbitrary potential $V (\zeta)$ of the three-form field. Thus, more degrees of freedom appear as compared to standard general relativity, in which all the properties of compact objects are determined only by baryons.

As a first step in our study, we have derived the basic equations describing the structure of static spherically symmetric compact objects in three-form field gravity, namely, the mass continuity equation, the generalized hydrostatic equilibrium equation (Tolman-Oppenheimer-Volkoff equation), and the field equation describing the evolution of the three-form field $\zeta$ in the given static geometry. An important physical parameter determining the properties of the stars in three-form field gravity is the self-interaction potential $V(\zeta)$ of the three-form field. In the present study we have adopted only two functional forms for $V(\zeta)$, by assuming that it is either constant inside the stars, or it is of the Higgs type, a second choice which is supported by the role such potentials play in elementary particle physics. The case of the constant potential can also be interpreted by assuming that the three-form field is in the minimum of the Higgs type potential, and therefore $V(\zeta)$ assumes a constant value.
Indeed, more general forms of the potential (exponential, hyperbolic, trigonometric, power-law etc) can also be considered, and we expect that the modification of the analytical form of the potential will have a significant impact on the stellar properties, leading to compact objects having different global properties.  Other choices of the potential would definitely generate different stellar sequences, corresponding to lighter or perhaps even more massive stars. Unfortunately, presently  there are no firm theoretical results establishing a clear relationship between the magnitude of the mass of the star and the field potential, and therefore the role of $V(A^2)$ can only be understood from the detailed numerical analysis of different stellar models under different physical and initial conditions.

Even in the framework of the simple  spherically symmetric static model the field equations of the three-form field theory become extremely complicated. In order to close the system of field equations we must specify the three-form field potential, as well as the equation of state of the baryonic matter. An alternative approach would be to specify the functional forms of the three-form field, together with the form of the baryonic matter density.  However, in our present study we have adopted a very general approach to the stellar structure problem, in which we have specified only the three-form field potential (as a constant, or Higgs type), and the equation of state of the baryonic matter.
We have considered four types of stellar models, corresponding to four choices of the baryonic matter equation of state.

The first of these equations, the stiff fluid equation of state \cite{Zeldovich:1962emp}, gives the limiting case of the causality condition, and guarantees that the speed of sound cannot exceed the speed of light in the baryonic matter. The radiation fluid equation of state plays an important role in astrophysics and physics, and it may be used to model the dense cores of neutron stars \cite{Misner:1964zz,R2}. Of particular interest are the quark matter \cite{Itoh:1970uw,Bodmer:1971we} and Bose-Einstein Condensate matter \cite{Boehmer:2007um,Harko:2019nyw} equations of state. If the quark model of hadrons is correct, then at high densities (not uncommon inside neutron stars) a deconfinement phase transition must take place, freeing the quarks from the bag, and leading to the formation of a quark star.  Quark stars may be born  during the collapse of the core of a massive neutron star after the supernova explosion, which may trigger a first or second order phase transition, by the conversion of neutron matter to quark matter in the core of a neutron star, or by the accretion of matter by neutron stars in low-mass X-ray binaries \cite{Glendenning:1997wn,Mak:2003kw}. On the other hand we may conjecture that in three-form field gravity a hadronic matter-quark matter phase transition can take place in extreme astrophysical and gravitational conditions, with the phase transition induced by the presence of the three-form field
with Higgs-type self-interaction potential. Some possible cosmic environments in which the phase transition could take place are gamma ray bursts, supernova explosions, or accretion by neutron stars. After such a phase transition
the three-form field star reaches a stable state in the minimum of the Higgs potential.

Superfluidity is also assumed to be a common occurrence in high density compact objects. The neutrons inside the star can form Cooper pairs, and a phase transition to a Bose-Einstein Condensate can also occur inside the star. Such a bosonic phase transition can also significantly affect the properties of compact objects \cite{Chavanis:2011cz}.

By numerically integrating the structure equations of the star, for the set of four equations of state we have considered, we have constructed four classes of three-form field gravity stellar models, corresponding to the stiff fluid, radiation fluid, quark matter and Bose-Einstein Condensate superfluid phase, respectively. In all of these cases we have thoroughly investigated the astrophysical properties of the stars (density, pressure and mass distributions, three-form field behavior), and compared them to the similar quantities obtained for standard general relativistic stars. In some of our numerical simulations the three-form potential attains negative values due to the specific choices of $V(A^2)$ and the relation between $A^2$ and $\zeta$, given by Eq.~\eqref{aux1}. It is also important to note that during our analysis, the specific cases where the three-form potential is given by a constant, the kinetic term of the form field, {\it i.e.} $F^2$ is also constant, rendering the three-form field energy density to behave in the same manner, that is $\rho_A=$cte. This is ascribed to a very well known fact \cite{Turok:1998he} that a massless three-form behaves exactly as a cosmological constant. In such cases, the theory reduces to pure GR plus a cosmological constant. Our investigations indicate that for all these four equations of state the three-form field gravity stars are generally more massive than their standard general relativistic counterparts. For example, for the stiff fluid equation of state, in the presence of the Higgs potential, three-form field
stars can reach maximum masses of the order of $5.2M_{\odot}$,  while the stiff fluid general relativistic stars may reach masses of the order of $3.2M_{\odot}$ \cite{Shapiro:1983du}. Three-form field stiff fluid stars in the presence of a constant potential could have masses as high as $4.2M_{\odot}$, which are much heavier than the general relativistic stars.

A similar trend can be detected in the case of photon stars, where in the presence of the Higgs-type potential the maximum mass of the star can reach values as high as $3M_{\odot}$. In the case of quark stars, whose maximum mass is of the order of $2M_{\odot}$ in standard general relativity, three-form field gravity can induce a significant increase of the mass, which, in the presence of the Higgs-type potential, can reach values as high as $2.7M_{\odot}$. Thus, for the same central density three-form field quark stars have significantly larger masses. The mass range of the considered  superfluid Bose-Einstein Condensate three-form field stars is around $1.97-2.08M_{\odot}$, which is still higher than the standard general relativistic mass of about $1.8M_{\odot}$. The mass of the three-form field star is also strongly influenced by its central density, with high central density stars having smaller gravitational masses.

Hence, it turns out that three-form field stars can have a very large mass spectrum. Recently, high precision determinations of the neutron star mass distribution have also confirmed the intriguing fact of the existence of neutron stars
with masses in the range of $2-2.20M_{\odot}$ \cite{Cromartie:2019kug, Horvath:2016akx,Demorest:2010bx,Antoniadis:2013pzd, vanKerkwijk:2010mt, Lattimer:2012nd,Ozel:2016oaf}. Such a neutron star having an unexpectedly high mass is the  eclipsing binary millisecond Black Widow Pulsar B1957+20, with the mass assumed to be in the range $1.6-2.4M_{\odot}$ \cite{vanKerkwijk:2010mt}. The recently discovered  J0740+6620 pulsar has an estimated mass of around $2.14M_{\odot}$ \cite{Cromartie:2019kug}. And of course there is the problem of the very high mass ($\sim 2.6M_{\odot}$) of the assumed neutron star in the GW190814 event \cite{Abbott:2020khf}.   A range of $2-2.4M_{\odot}$ is very difficult to explain by the standard hadronic matter models, constructed by using the present day knowledge of the nuclear equations of state, and in the framework of general relativity. Even the consideration of exotic models like quark or kaon stars cannot lead to a better understanding of the observations.

While explaining this mass range is problematic in standard general relativity, it can be easily explained in three-form field gravity, where one can construct stellar models having these numerical values of the masses, by using the standard knowledge about the equation of state of dense matter.
The large mass spectrum of the three-form field gravity stars also could point towards the possibility
that high mass objects, usually interpreted as stellar mass black holes, and having masses in the range of $3.8M_{\odot}$ and $6M_{\odot}$, respectively, could be in fact three-form field stars.  Such a possibility has been already investigated in \cite{Kovacs:2009kv} for the case of the quarks stars in the Color-Flavor Locked phase.  On the other hand, based on the present results on the structure and properties of three-form field stars,  the possibility that at least some stellar mass black holes are in fact ordinary three-form field stars in the presence of a field potential cannot be rejected a priori. As we have shown in detail, three-form field gravity
stars could have much higher masses than standard neutron stars, and thus they represent some ordinary star alternatives to low mass  black hole candidates.

A basic problem in theoretical astrophysics is finding a convincing way that could differentiate between different types of stellar models, and which also put some constraints on the equation of state of the high density matter.  One of these possibilities may be related to the determination of the surface redshift of different types of compact objects. The surface redshift is defined as
\be
z=\left(1-\frac{2GM}{c^2R}\right)^{-1/2}-1.
\ee

In standard general relativity, where the mass-radius ratio satisfies the Buchdahl inequality $2GM/c^2R\leq 8/9$ \cite{Buchdahl}, the surface redshift satisfies the constraint $z\leq 2$. A selected sample of surface redshifts of the maximum mass stars of all considered three-form field models is presented in Table~\ref{table9}, with the corresponding standard general relativistic values also presented. As one can see from the Table, there are significant differences between the general relativistic and the three-form field stars from the point of view of the redshift. Important differences also appear between different types of stars, indicating the important role the equation of state of dense matter plays in the description of the global astrophysical parameters of relativistic stars. Negative values of the constant three-form potential lead to smaller redshift values as compared to the GR case, while positive values of $\lambda$, as well as the Higgs potential case lead to higher redshift values. An interesting case is represented by the Bose-Einstein Condensate three-form field stars, having almost the same surface redshift as their general relativistic counterparts. However, distinguishing different types of stellar objects based solely on their surface redshift may prove to be an extremely difficult observational task, which would require a significant increase in the observational accuracy.

\begin{center}
	\begin{table*}[htbp]
		\begin{tabular}{|c|c|c|}
				\hline
				\textbf {Stars}&\textbf{Model parameters}& \textbf{Surface Redshift}\\
				\hline
				\multirow{7}{*}{\textbf{Stiff fluid stars}}& GR ($\bar{\zeta}=0$)&$0.45$
				\\	
				&$\bar{\lambda}=-0.02,~~\bar{a}=\bar{b}=0.0$&$0.37$\\
				&$\bar{\lambda}=\bar{a}=\bar{b}=0.0$&$0.49$\\
				&$\bar{\lambda}=0.02,~~\bar{a}=\bar{b}=0.0$&$0.66$\\
				&~~$\bar{\lambda}=0.0,~~\bar{a}=0.003,~~\bar{b}=-0.033$&$~1.09$~~~\\
				&$\bar{\lambda}=0.0,~~\bar{a}=0.003,~~\bar{b}=0.0$&$0.78$\\
				&~~$\bar{\lambda}=0.0,~~\bar{a}=0.003,~~\bar{b}=0.033$&$0.67$~\\
				\hline
				\hline
				\multirow{7}{*}{\textbf{Photon stars}}& GR ($\bar{\zeta}=0$)&$0.37$
				\\
				&$\bar{\lambda}=-0.02,~~\bar{a}=\bar{b}=0.0$&$0.34$\\
				&$\bar{\lambda}=\bar{a}=\bar{b}=0.0$&$0.39$\\
				&$\bar{\lambda}=0.02,~~\bar{a}=\bar{b}=0.0$&$0.47$\\
				&~~$\bar{\lambda}=0.0,~~\bar{a}=0.007,~~\bar{b}=-0.11$&$0.67$~\\
				&$\bar{\lambda}=0.0,~~\bar{a}=0.007,~~\bar{b}=0.0$&$0.54$\\
				&~~$\bar{\lambda}=0.0,~~\bar{a}=0.007,~~\bar{b}=0.11$&$0.49$~\\
				\hline
				\hline
				\multirow{7}{*}{\textbf{Quark stars}}& GR ($\bar{\zeta}=0$)&$0.48$
				\\
				&$\bar{\lambda}=-0.02,~~\bar{a}=\bar{b}=0.0$&$0.44$\\
				&$\bar{\lambda}=\bar{a}=\bar{b}=0.0$&$0.50$\\
				&$\bar{\lambda}=0.02,~~\bar{a}=\bar{b}=0.0$&$0.57$\\
				&~~$\bar{\lambda}=0.0,~~\bar{a}=0.01,~~\bar{b}=-0.11$&$0.75$~\\
				&$\bar{\lambda}=0.0,~~\bar{a}=0.01,~~\bar{b}=0.0$&$0.67$\\
				&~~$\bar{\lambda}=0.0,~~\bar{a}=0.01,~~\bar{b}=0.11$&$0.62$~\\
				\hline
				\hline
				\multirow{7}{*}{\textbf{Bose-Einstein Condensate stars}}& GR ($\bar{\zeta}=0$)&$0.46$
				\\
				&$\bar{\lambda}=-0.01,~~\bar{a}=\bar{b}=0.0$&$0.46$\\
				&$\bar{\lambda}=\bar{a}=\bar{b}=0.0$&$0.45$\\
				&$\bar{\lambda}=0.0013,~~\bar{a}=\bar{b}=0.0$&$0.44$\\
				&~~$\bar{\lambda}=0.0,~~\bar{a}=0.001,~~\bar{b}=-0.1$&$0.46$~\\
				&$\bar{\lambda}=0.0,~~\bar{a}=0.001,~~\bar{b}=0.0$&$0.46$\\
				&~~$\bar{\lambda}=0.0,~~\bar{a}=0.001,~~\bar{b}=0.6$&$0.45$~\\
				\hline
				\end{tabular}
	\caption{The surface redshift of different types of three-form field stars with maximum mass, for different model parameters.}
			\label{table9}
				\end{table*}
	\end{center}

Another important observational possibility that may allow to distinguish between three-form field  stars and standard general relativistic stars or
stellar mass black holes could be represented by the detailed study of the physical and astrophysical properties of the thin accretion disks that form around compact objects. The radiation properties of the accretion disks around general relativistic stars and black holes and three-form field stars are different due to the differences in both internal and external geometry. Therefore the electromagnetic emission properties of the accretion disk (energy flux, temperature, luminosity), and of the compact central objects, may provide the basic signature that could lead to the convincing differentiation between modified gravity stars and black holes from their standard general relativistic counterparts
\cite{Harko:2009gc,Harko:2009rp,Harko:2009kj,Harko:2010ua,Danila:2015qla,Shahidi:2020bla}. Multimessenger Astronomy and gravitational wave observations will also certainly play an important role in this direction.

Three-form field gravity stars manifest a very complex internal structure, which also determines a complex stellar dynamics. The presence of the three-form field can lead to a number of distinctive astrophysical signatures, which could help in the observational detection of these types of objects. However, this may prove to be an extremely difficult task. Further issues related to the astrophysical/observational importance of the three-form field gravity stars will be considered in a future publication.\\

\section*{Acknowledgements}
We would like to thank to anonymous reviewer for comments and suggestions that helped us to improve our manuscript. BJB is supported by the grant PD/BD/128018/2016 (PhD::SPACE program). FSNL acknowledges support from the Funda\c{c}\~{a}o para a Ci\^{e}ncia e a Tecnologia (FCT) Scientific Employment Stimulus contract with reference CEECIND/04057/2017 BJB and FSNL also acknowledge funding from the FCT research grants No. UID/FIS/04434/2020, No. PTDC/FIS-OUT/29048/2017 and No. CERN/FIS-PAR/0037/2019.



\end{document}